\documentclass[draftcls, onecolumn, 11pt]{IEEEtran}
\usepackage{amsmath,amssymb,threeparttable,placeins,enumerate,caption}
\usepackage{mathtools,relsize}
\usepackage[pdftex]{graphicx}
\usepackage{epstopdf}
\usepackage{color}
\usepackage[usenames]{xcolor}
\usepackage{amsfonts}
\usepackage{latexsym}
\usepackage{subfigure,multirow,makecell,stfloats}
\usepackage{lscape}
\usepackage[figuresright]{rotating}
 \usepackage{array}
\usepackage{algorithm}
\usepackage{cite}
\usepackage[noend]{algpseudocode}

\def\qi#1 {\fbox {\footnote {\ }}\ \footnotetext { From Qi: {\color{red}#1}}}
\usepackage{amssymb}

\usepackage{amsthm}

\theoremstyle{remark} % 编号显示风格，可选 plain or definition or remark
\newtheorem{theorem}{{{\textit{Theorem}}}}

\newtheorem{lemma}{{{\textit{Lemma}}}}
\newtheorem{corollary}{{{{\textit{Corollary}}}}}

\newtheorem{definition}{{{\textit{Definition}}}}

\newtheorem{remark}{{{\textit{Remark}}}}

\newtheorem{example}{{{\textit{Example}}}}

\newtheorem{construction}{{{\textit{Construction}}}}
\title{Asymptotically Optimal Aperiodic and Periodic Sequence Sets with Low Ambiguity Zone Through Locally Perfect Nonlinear Functions}

	\author{Zheng Wang, Zhengchun Zhou,~\IEEEmembership{Member,~IEEE}, Avik Ranjan Adhikary,~\IEEEmembership{Member,~IEEE},
	Yang Yang,~\IEEEmembership{Member,~IEEE}, Sihem Mesnager, and Pingzhi Fan,~\IEEEmembership{Fellow,~IEEE}.
	\thanks{ 
		 Z. Wang, A. R. Adhikary, and Y. Yang are with the School of Mathematics, Southwest Jiaotong University, Chengdu, 611756, China. (e-mail: wang\_z@my.swjtu.edu.cn, avik.adhikary@ieee.org, yang\_data@qq.com).
		
		 Z. Zhou is with the School of Information Science and Technology, Southwest Jiaotong University, Chengdu, 611756, China. (e-mail: zzc@swjtu.edu.cn).
		 
		S. Mesnager is with the Department of Mathematics, University of
		Paris VIII, 93526 Saint-Denis, France, also with CNRS, UMR 7539 LAGA,
		University of Paris XIII, 93430 Villetaneuse, France (e-mail:
		smesnager@univ-paris8.fr).
		
		P. Fan is with the Institute of Mobile Communications, Southwest
		Jiaotong University, Chengdu 611756, China (e-mail: p.fan@ieee.org).
	}
}

\begin{document}
	\maketitle
	\begin{center}
		\textbf{Abstract}
	\end{center}

Low ambiguity zone (LAZ) sequences play a crucial role in modern integrated sensing and communication (ISAC) systems. In this paper, we introduce a novel class of functions known as locally perfect nonlinear functions (LPNFs). By utilizing LPNFs and interleaving techniques, we propose three new classes of both periodic and aperiodic LAZ sequence sets with flexible parameters. The proposed periodic LAZ sequence sets are asymptotically optimal in relation to the periodic Ye-Zhou-Liu-Fan-Lei-Tang bound. Notably, the aperiodic LAZ sequence sets also asymptotically satisfy the aperiodic Ye-Zhou-Liu-Fan-Lei-Tang bound, marking the first construction in the literature. Finally, we demonstrate that the proposed sequence sets are cyclically distinct.
	
	\begin{center}
		\textbf{Index Terms}
	\end{center}
	
	Ambiguity function (AF), Low Ambiguity Zone (LAZ),  (Locally) Perfect Nonlinear Function, Sequence, Joint communication and sensing (JCAS), Integrated Sensing and Communication (ISAC) system.

	\section{Introduction}
	
Joint communication and sensing (JCAS) is an emerging technology with significant potential that has garnered substantial attention in recent years. The NASA Space Shuttle Orbiter  (\cite{nasa}) is recognized as the first known application of an Integrated Sensing and Communication (ISAC) system for space missions. In military contexts, the Advanced Multifunction Radio Frequency Concept (AMRFC) addresses the need to consolidate radar, communications, and electronic warfare functions into a single platform. Additionally, a standardization body has begun preliminary efforts toward the standardization of ISAC  \cite{3gpp,3gpp1}.
According to a technical report released in 2024  (\cite{3gpp,3gpp1}) by the mentioned standardization body, ISAC has the potential to play a crucial role in various applications. These include the detection and tracking of unmanned aerial vehicles (UAVs), smart transportation systems, navigation assistance, collision avoidance for automated guided vehicles (AGVs) and autonomous mobile robots (AMRs), as well as enhancements to millimeter-wave (mmWave) communications. The term ``integration" in ISAC can refer to different levels of integration, wherein communication and sensing share the same sites, spectrum, or hardware \cite{Liu22,Liu22_1}. The highest level of integration involves reusing signals transmitted for communication purposes. While this approach can lead to reduced overhead and energy consumption, it is effective only in scenarios where the communication signal already has the capability to perform sensing functions \cite{Liu22,Liu22_1}.

Sequences play a significant role in designing signals with various capabilities \cite{SD2005}. The Ambiguity Function (AF) serves as a metric to evaluate the communication and sensing capabilities of these sequences \cite{SD2005}. For Integrated Sensing and Communication (ISAC) systems, sequences with zero or low ambiguity sidelobes across the delay-Doppler domain are preferred \cite{hassanien2019dual}. 

In 2013, Ding and Feng \textit{et al.} \cite{ding13} established the theoretical lower bound for the maximum sidelobe of the AF based on the Welch bound. They also introduced sequences with favorable AF properties, including a single sequence that achieves this theoretical bound throughout the delay-Doppler (DD) region \cite{ding13}. Furthermore, Wang and Gong developed several classes of sequences characterized by low AF magnitudes across the entire DD region, utilizing the multiplication and addition properties in finite fields \cite{gong13,wang11,wang13}. Achieving low AF sidelobes across the entire DD region is generally challenging and must satisfy the criteria established in \cite{ding13}.

There are also a few constructions that employ optimization algorithms. In \cite{soltanalian2014designing}, the authors proposed an algorithm based on a monotonically error-bound improving technique (MERIT) to search for low aperiodic AF sidelobes within a specific DD region. An accelerated iterative sequential optimization (AISO) algorithm was introduced in \cite{CD2017} to enhance convergence rates and reduce computational complexity, based on a general framework of majorization-minimization (MM) algorithms \cite{mmstoica2004cyclic}, \cite{mmrazaviyayn2013unified}. Moreover, additional studies \cite{He2012}, \cite{liu2019unimodular}, \cite{wang2022unified} also focused on finding sequences with low AF sidelobes.

However, optimization algorithms become impractical for designing sequences as their complexity and length increase. The challenge of systematically constructing sequences with zero or low AF sidelobes that achieve the upper bound proposed in \cite{ding13} with equality across the entire DD region remains unresolved to this day.

It is important to note that in many practical applications (for example, in \cite{duggal2020doppler, kumari}), the Doppler frequency range is often significantly smaller than the bandwidth of the transmitted signal. Consequently, it may not be necessary to consider the entire duration of the signal.  In this context, Ye \textit{et al.} \cite{ye22} made significant progress by introducing sequences known as zero/low ambiguity zone sequences (ZAZ/LAZ sequences), which exhibit zero or low sidelobes only within a specific region around the origin of the delay-Doppler (DD) region. The authors proposed bounds on the size of the zone and the maximum magnitude of the ambiguity function (AF) sidelobes for ZAZ/LAZ sequences within the low ambiguity zone (LAZ). These bounds are referred to as the  ``\textit{Ye-Zhou-Liu-Fan-Lei-Tang bounds}".
Additionally, they introduced optimal unimodular ZAZ/LAZ sequences by utilizing certain cubic sequence families and Zadoff-Chu sequences. In \cite{cao2022interleaved}, Cao \textit{et al.} observed that a class of binary linear constrained zero (LCZ) sequence sets presented in \cite{zhou2008new} exhibits low ambiguity properties within a DD region around the origin. Furthermore, Tian \textit{et al.} \cite{tian2024asymptotically} proposed a set of sequences with optimal ZAZ/LAZ properties by employing perfect nonlinear functions (PNFs). Recently, Meng \textit{et al.} \cite{meng2024new} proposed a new aperiodic bound that is tighter than the aperiodic Ye-Zhou-Liu-Fan-Lei-Tang bound for specific parameters of LAZ sequences. The authors also suggested an order-optimal aperiodic LAZ sequence set related to that bound.

Motivated by the works of \cite{ye22} and \cite{tian2024asymptotically}, this work focuses on designing LAZ sequence sets with more flexible parameters. We recognize that perfect nonlinear functions (PNFs) \cite{pnfs} play a significant role in the design of LAZ sequences in \cite{ye22} and \cite{tian2024asymptotically}. However, the limited availability of PNF makes it challenging to design LAZ sequences with greater flexibility. To address this issue, we propose a new class of functions called locally perfect nonlinear functions (LPNFs), which exhibit perfect nonlinearity within a specific region.

Using LPNFs and interleaving techniques, we introduce a new framework for designing LAZ sequence sets with flexible parameters. We create three classes of periodic and aperiodic LAZ sequence sets using this framework. By selecting appropriate parameter values, we demonstrate that the proposed periodic and aperiodic LAZ sequence sets are asymptotically optimal with respect to the periodic and aperiodic AF bounds proposed in \cite{ye22}. The proposed aperiodic LAZ sequence sets are the first to asymptotically satisfy the aperiodic Ye-Zhou-Liu-Fan-Lei-Tang bound. We also show that the proposed sequences are cyclically distinct. Notably, the LAZ sequence sets proposed in \cite{tian2024asymptotically} can be regarded as a special case of our construction when the parameters are appropriately selected. Table \ref{t1} lists the parameters of ZAZ/LAZ sequences proposed to date.

%\begin{landscape}
\begin{sidewaystable}
	%	\begin{table}[htp]
	\caption{Comparison of unimodular LAZ/ZAZ sequence sets}\label{t1}
	\resizebox{\linewidth}{!}{
	%	\resizebox{23cm}{!}{
		\renewcommand{\arraystretch}{1.1}
		\begin{tabular}{|c|c|c|c|c|c|c|c|c|c|}
			\hline \multicolumn{2}{|c|}{ Method } & Length & Set size & $\theta_{\max }(\hat{\theta}_{\max})$ & $Z_x$ & $Z_y$ & Constraints & Optimality & Remarks\\
			\hline \multirow{4}{*}{ \cite{ding13} } & \it{Theorem 9} & $L=q-1$ & 1 & $\sqrt{q}$ & $L$ & $L$ & $q=p^l, p$ is a prime & optimal & Periodic\\
			\cline{2-10} & \it{Theorem 10} & $L=\frac{q-1}{N}$ & $N$ & $\sqrt{q}$ & $L$ & $L$ & $q=p^l, p$ is a prime, $N \mid(q-1), N \geq 2$ & & Periodic\\
			\cline{2-10} & \multirow[b]{2}{*}{ \it{Theorem 11}} & $L=\frac{q-1}{N}$ & $L$ & $2 \sqrt{q}$ & $L$ & $L$ & $q=p^l, p$ is a prime, $N \mid(q-1), N \geq 2$ & & Periodic\\
			\cline{3-10} & & $L=\frac{q-1}{N}$ & $\prod_{j=1}^s\left(\frac{L}{\left(L \cdot \lambda_i\right)}+1\right)$ & $\frac{\lambda_s \sqrt{9}}{L}$ & $L$ & $L$ & \begin{tabular}{l}
				$q=p^l, p$ is a prime, $N \mid(q-1), N \geq 2,\left\{\lambda_i\right\}_{i=0}^{\infty}$ \\
				are positive integers coprime to $q$ in increasing order
			\end{tabular} & & Periodic\\
			\hline \multirow{4}{*}{ \cite{ye22} } & \it{Theorem 5} & $p$ & $p^2$ & $2 \sqrt{p}$ & $p$ & $p$ & $p$ is an odd prime & & Periodic\\
			\cline{2-10} & \it{Theorem 6} & $p$ & 1 & $\sqrt{p}$ & $p$ & $p$ & $p$ is an odd prime & optimal & Periodic\\
			\cline{2-10} & \it{Theorem 8} & $L$ & 1 & 0 & $\frac{L}{r}$ & $r$ & $\operatorname{gcd}(a, L)=1$ if $L$ is odd, $r=\operatorname{gcd}(2 a, L), r>1$ & optimal & Periodic\\
			\cline{2-10} & \it{Construction 2} & $L$ & $N$ & 0 & $\frac{\mid L / N\rfloor}{r}$ & $r$ & $\operatorname{gcd}(a, L)=1$ if $L$ is odd, $r=\operatorname{gcd}(2 a, L), r>1$ & optimal if $N \mid L$ & Periodic\\
			\hline \cite{cao2022interleaved} & \it{Theorem 4} & $2\left(2^m-1\right)$ & $2 M$ & $2 \sqrt{2^m}$ & $L$ & $2\left(2^m-1\right)$ & \begin{tabular}{l}
				\begin{tabular}{c}
					$E=\left\{e_i=\left(e_{i, 0}, e_{i, 1}\right): 0 \leq i<M\right\}$ is a shift sequence \\
					set, $L=\min \left\{\min _{e \neq h \in E}\left\{2\left(e_0-h_0\right), 2\left(e_1-h_1\right)\right\}\right.$, \\
					$\left.\min _{e \neq h \in E}\left\{2\left(e_0-h_1\right)+1,2\left(e_1-h_0\right)-1\right\}\right\}$
				\end{tabular}
			\end{tabular} & & Periodic\\
			\hline
			\cite{meng2024new} & \it{Corollary 8} & $N$ & $M$ & $3C\sqrt{N}$ & $\left\lfloor\frac{N}{\max\limits_{m\in[1,M]}\left|a_m\right|}-1\right\rfloor$ & $\min\limits_{m\in[1,M]}\left|a_m\right|+1$ & \begin{tabular}{l}
				\begin{tabular}{c}
					$C=\max\limits_{{m_1},{m_2}\in[1,M-1]} \left\{\sqrt{|a_{m_1}-a_{m_2}|}+\frac{2}{\sqrt{|a_{m_1}-a_{m_2}|}}\right\},$\\
					$N \geq 5 \max\limits_{m\in[1,M]}\left|a_m\right|$ and $\max\limits_{m_1, m_2 \in[1, M]}\left\{a_{m_1}-a_{m_2}\right\} \ll N$,\\
					$a_{m}\in[1,N-1], m\in[1,M]$
				\end{tabular}
			\end{tabular} &  & Aperiodic\\
			\hline \multirow{3}{*}{ \cite{tian2024asymptotically} } & \it{Corollary 1} & $M N^2$ & $M N$ & 0 & $\left\lfloor\frac{N}{K}\right\rfloor$ & $K$ & $K<N, \operatorname{gcd}(K, N)=1$ & asymptotically optimal & Periodic\\
			\cline{2-10} & \it{Theorem 3} & $N(K N+P)$ & $N$ & 0 & $N$ & $K$ & $\operatorname{gcd}(P, N K)=1$ & asymptotically optimal & Periodic\\
			\cline{2-10} & \it{Theorem 4} & $p(p-1)$ & $p$ & $p$ & $p-1$ & $p$ & $p$ is an odd prime & asymptotically optimal & Periodic\\	
			\hline \multirow{6}{*}{ This paper } & 
			\it{Corollary \ref{Th3}} & $N^2$ & $N$ & $N$ & $p$ & $N$ & $N$ is an odd number, $p$ is the smallest prime factor of $N$  & asymptotically optimal & Periodic\\
			\cline{2-10} & \it{Corollary \ref{Th3}} & $NK$ & $N$ & $K$ & $p$ & $K-N+1$ &  
			\begin{tabular}{l}
				\begin{tabular}{c}
					$N$ is an odd number, $p$ is the smallest prime factor of $N$,\\
					$N<K<2N-1$ 
				\end{tabular}
			\end{tabular}& asymptotically optimal  & Periodic\\
			\cline{2-10} & \it{Corollary \ref{Th3}} & $NK$ & $N$ & $K$ & $p$ & $K$ & \begin{tabular}{l}
				\begin{tabular}{c}
					$N$ is an odd number, $p$ is the smallest prime factor of $N$,\\
					$K\geq 2N-1$ 
				\end{tabular}
			\end{tabular} &  & Periodic\\
			\cline{2-10} &  \it{Corollary \ref{Th4}} & $N^2$ & $N$ & $N+p-1$ & $p$ & $N$ & $N$ is an odd number, $p$ is the smallest prime factor of $N$ & asymptotically optimal & Aperiodic\\
			\cline{2-10} &  \it{Corollary \ref{Th4}} & $NK$ & $N$ & $K+p-1$ & $p$ & $K-N+1$ & 	\begin{tabular}{l}
				\begin{tabular}{c}
					$N$ is an odd number, $p$ is the smallest prime factor of $N$,\\
					$N<K<2N-1$ 
				\end{tabular}
			\end{tabular} & asymptotically optimal & Aperiodic\\
			\cline{2-10} &  \it{Corollary \ref{Th4}} & $NK$ & $N$ & $K+p-1$ & $p$ & $K$ & \begin{tabular}{l}
				\begin{tabular}{c}
					$N$ is an odd number, $p$ is the smallest prime factor of $N$,\\
					$K\geq 2N-1$ 
				\end{tabular}
			\end{tabular}  && Aperiodic\\
			\hline
	\end{tabular}}
	\\
	$\theta_{\max}(\hat{\theta}_{\max })$ is the maximum periodic(aperiodic) AF magnitude for $(\tau, v) \in\left(-Z_x, Z_x\right) \times\left(-Z_y, Z_y\right)$, where $\tau$ is time delay and $v$ is Doppler shift.
	%	\end{table}
\end{sidewaystable}
%\end{landscape}	
	
The rest of this paper is organized as follows. Section \ref{preliminary} introduces the necessary notations and concepts, discussing the prerequisites for the subsequent sections. Next, we present various concepts and develop our methodology's framework for constructing periodic and aperiodic LAZ sequence sets. Specifically, Section \ref{section-LPNF} discusses ``locally" perfect nonlinear functions, a concept derived from perfect nonlinear functions we introduced earlier. This concept will be a crucial component in establishing our construction of sequence sets. Section \ref{Constructions} focuses on sequence designs, specifically presenting our constructions of derived sequence sets. We also evaluate the quality of the produced LAZ sequence sets in terms of optimality, demonstrating that these sets are asymptotically optimal. Finally, Section \ref{conclusion} concludes the paper.

	\section{preliminaries}\label{preliminary}
	For convenience, we will use the following notations consistently in this paper:
	\begin{itemize}
		\item $\mathbb{Z}_N$ denotes the ring of integers modulo $N$.
		\item $\mathbb{Z}_N^{*}$ denotes the multiplcative group of ring $\mathbb{Z}_N.$ 
		\item $\omega_N=e^{\frac{2 \pi \sqrt{-1}}{N}}=e^{\frac{2 \pi i}{N}}$ is a primitive $N$-th complex root of unity.
		%	\item $\lfloor a / b\rfloor$ denotes the largest integer not greater than $a / b$.
		\item $\mathcal{S}$ denotes a sequence set.
		\item $\mathbf{s}$ denotes a sequence.
		\item $\mathbf{s}^T$ denotes the transpose of sequence $\mathbf{s}$.
		\item $x^*$ denotes the complex conjugate of $x$. 
		%		\item $\langle x \rangle_N$ denotes $x \pmod N$.
	\end{itemize}

	\subsection{Ambiguity Function}

	Let $\mathbf{a}=(a(0),a(1),\cdots,a(N-1))$ and $\mathbf{b}=(b(0),b(1),$ $\cdots,b(N-1))$ be two unimodular sequences with period $N.$ 
	The aperiodic cross AF of $\mathbf{a}$ and $\mathbf{b}$ at time shift $\tau$ and Doppler $v$ is defined as follows:

\begin{align}
		\hat{A F}_{\mathbf{a}, \mathbf{b}}(\tau, \nu)=\left\{\begin{aligned}
			&\sum_{t=0}^{N-1-\tau} a(t) b^*(t+\tau) \omega_N^{vt}, & 0 \leq \tau \leq N-1,\\
			&\sum_{t=-\tau}^{N-1} a(t) b^*(t+\tau) \omega_N^{vt}, & N \leq \tau<0,\\
			&0,&|\tau|\geq N.
		\end{aligned}\right.
	\end{align}
	
We define the periodic cross-ambiguity function as follows:
\begin{align*} 
AF_{\mathbf{a}, \mathbf{b}}(\tau, v) = \sum_{t=0}^{N-1} a(t) b^*(\langle t+\tau \rangle_N) \omega_N^{vt},
\end{align*}

where \(\langle \cdot \rangle_N\) denotes the modulo operation with respect to \(N\). If \(\mathbf{a} = \mathbf{b}\), then \(AF_{\mathbf{a}, \mathbf{b}}(\tau, v)\) is called the periodic auto-AF, which is denoted as \(AF_{\mathbf{a}}(\tau, v)\).

\subsection{Low Ambiguity Zone}

	For a sequence set $\mathcal{S}$ of size $M$ with length $N$, its maximum periodic AF magnitude over a region $\Pi =(-Z_x, Z_x) \times(-Z_y, Z_y)\subseteq(-N, N) \times(-N, N)$ is defined as $\theta_{\max}(\mathcal{S})=\max \left\{\theta_A(\mathcal{S}), \theta_C(\mathcal{S})\right\}$, the maximum periodic auto AF magnitude
	\begin{align*}
		\theta_A(\mathcal{S})=\max \left\{\left|A F_{\mathbf{a}}\left(\tau, v\right)\right|: \mathbf{a} \in \mathcal{S},(0,0) \neq \left(\tau,v\right) \in \Pi\right\},
	\end{align*}
	and the maximum periodic cross AF magnitude
	\begin{align*}
		\theta_C(\mathcal{S})=\max \left\{\left|A F_{\mathbf{a}, \mathbf{b}}\left(\tau, v\right)\right|: \mathbf{a} \neq \mathbf{b} \in \mathcal{S}, \left(\tau,v\right) \in \Pi\right\} .
	\end{align*}
	
	The sequence set  $\mathcal{S}$ is referred to as an $(M, N, \Pi,\theta_{\max})$-LAZ periodic sequence set, where $M$ denotes the set size, $N$ denotes the sequence length, $\Pi$ denotes the low ambiguity zone, and $\theta_{\max}$ denotes the maximum periodic AF magnitude in region $\Pi.$
	
	Its maximum aperiodic AF magnitude over a region $\Pi =(-Z_x, Z_x) \times(-Z_y, Z_y)\subseteq(-N, N) \times(-N, N)$ is defined as $\hat{\theta}_{\max}(\mathcal{S})=\max \left\{\hat{\theta}_A(\mathcal{S}), \hat{\theta}_C(\mathcal{S})\right\}$, the maximum aperiodic auto AF magnitude
	\begin{align*}
		\hat{\theta}_A(\mathcal{S})=\max \left\{\left|\hat{AF}_{\mathbf{a}}\left(\tau, v\right)\right|: \mathbf{a} \in \mathcal{S},(0,0) \neq \left(\tau,v\right) \in \Pi\right\},
	\end{align*}
	and the maximum aperiodic cross AF magnitude
	\begin{align*}
		\hat{\theta}_C(\mathcal{S})=\max \left\{\left|\hat{AF}_{\mathbf{a}, \mathbf{b}}\left(\tau, v\right)\right|: \mathbf{a} \neq \mathbf{b} \in \mathcal{S}, \left(\tau,v\right) \in \Pi\right\}.
	\end{align*}
	
	The  sequence set $\mathcal{S}$ is referred to as an $(M, N, \Pi,\hat{\theta}_{\max})$-LAZ aperiodic sequence set, where $M$ denotes the set size, $N$ denotes the sequence length, $\Pi$ denotes the low ambiguity zone, and $\hat{\theta}_{\max}$ denotes the maximum aperiodic AF magnitude in region $\Pi.$

	\subsection{Bounds of LAZ Sequences Set and Optimality Factor}
	\begin{lemma}\label{PLAZ}
    (\cite{ye22}) For any unimodular sequence set $\mathcal{S}= (M, N, \Pi, \theta_{\max})$, where $\Pi = (-Z_x, Z_x) \times (-Z_y, Z_y),$ the maximum periodic AF magnitude satisfies the following lower bound:
    $$
    \theta_{\max} \geq \frac{N}{\sqrt{Z_y}} \sqrt{\frac{M Z_x Z_y / N - 1}{M Z_x - 1}}.
    $$
\end{lemma}

To evaluate the closeness between the theoretical lower bound and the achieved lower bound of the periodic AF magnitude of the sequence sets proposed through systematic constructions, we define the optimality factor $\rho_{\mathrm{LAZ}}$ as follows. 
\begin{definition}
	For any unimodular sequence set $\mathcal{S}= (M, N, \Pi, \theta_{\max})$, where $\Pi = (-Z_x, Z_x) \times (-Z_y, Z_y),$ the optimality factor $\rho_{\mathrm{LAZ}}$ is defined as follows:
$$
\rho_{\mathrm{LAZ}} = \frac{\theta_{\max}}{\frac{N}{\sqrt{Z_y}} \sqrt{\frac{M Z_x Z_y / N - 1}{M Z_x - 1}}}.
$$
\end{definition}
In general, $\rho_{\mathrm{LAZ}} \geq 1$. The LAZ sequence set is said to be optimal if the ratio of the achieved maximum periodic AF magnitude within the LAZ and the theoretical bound is equal to 1, i.e., $\rho_{\mathrm{LAZ}}=1$. When $\lim\limits_{N \rightarrow \infty}\rho_{\mathrm{LAZ}} = 1$, the LAZ sequence set is considered as asymptotically optimal.

\begin{lemma}\label{APLAZ}(\cite{ye22}) 
For any unimodular sequence set \(\mathcal{S} = (M, N, \Pi, \hat{\theta}_{\max})\), where $\Pi = (-Z_x, Z_x) \times (-Z_y, Z_y)$, the maximum aperiodic AF magnitude satisfies the following lower bound:
    $$
    \hat{\theta}_{\max} \geq \frac{N}{\sqrt{Z_y}} \sqrt{\frac{M Z_x Z_y - N - Z_x + 1}{(N + Z_x - 1)(M Z_x - 1)}}.
    $$
\end{lemma}

Similar to $\rho_{\mathrm{LAZ}}$ we define the optimality factor $\hat{\rho}_{\mathrm{LAZ}}$ to evaluate the closeness between the theoretical lower bound and the achieved lower bound of the aperiodic AF magnitude of the sequence sets proposed through systematic constructions.
\begin{definition}
	For any unimodular sequence set $\mathcal{S}= (M, N, \Pi, \hat{\theta}_{\max})$, where $\Pi = (-Z_x, Z_x) \times (-Z_y, Z_y),$ the optimality factor $\rho_{\mathrm{LAZ}}$ as follows:
$$
\hat{\rho}_{\mathrm{LAZ}} = \frac{\hat{\theta}_{\max}}{\frac{N}{\sqrt{Z_y}} \sqrt{\frac{M Z_x Z_y - N - Z_x + 1}{(N + Z_x - 1)(M Z_x - 1)}}}.
$$
\end{definition}
The LAZ sequence set is said to be optimal if the ratio of the achieved maximum aperiodic AF magnitude within the LAZ and the theoretical bound is equal to 1, i.e., $\hat{\rho}_{\mathrm{LAZ}}=1$. When $\lim\limits_{N \rightarrow \infty}\hat{\rho}_{\mathrm{LAZ}} = 1$, the LAZ sequence set is considered as asymptotically optimal.
	
	\subsection{Perfect Nonlinear Functions and Related Concepts}

\begin{definition}\label{PNFs}
Let \( A \) and \( B \) be two finite Abelian groups, with their cardinalities denoted by \(|A|\) and \(|B|\), respectively. Let \( f \) be a function from \( A \) to \( B \). The measure of nonlinearity of \( f \) is defined as follows:

\[
P_f = \max_{0 \neq a \in A} \max_{b \in B} | \{ x \in A : f(x + a) - f(x) = b \} |.
\]

If the quantity \( P_f \) is equal to \( \frac{|B|}{|A|} \), then \( f \) is called a perfect nonlinear function (PNF).
    \end{definition}

Readers interested in further information about PNFs can go through \cite{pnfs}(Chapter 3) and the references therein.

When \( f \) is a mapping between two Abelian groups \( A \) and \( B \) of the same size, we can define a difference map for any element \( a \in A \) where \( a \neq 0 \). This is given by \( \Delta_{f,a}(x) = f(x + a) - f(x) \). This difference map measures the degree of ``linearity" of \( f \). A function \( f \) is termed perfect nonlinear if \( \Delta_{f,a} \) is injective, while it is called almost perfect nonlinear (APN) if \( \Delta_{f,a} \) is at most $2$-to-$1$. These functions are of significant interest due to their resistance to linear and differential cryptanalysis. Notably, if \( f \) is bijective, there will never be solutions for \( b = 0 \). In such cases, the relevant notion corresponds to \( b \) being restricted to \( B \setminus \{0\} \).The concept of ambiguity in this context has been more thoroughly studied, particularly in \cite{Panario2011}, which focuses on the ambiguity of permutations between finite Abelian groups of the same size.  In this paper, we will concentrate solely on the PNFs that we have revised within our framework in the next Section \ref{section-LPNF}.

	\subsection{On the Interleaving Technique}
	
The interleaving technique has been a vital tool in sequence design since 1995, when Gong introduced it for constructing longer sequences \cite{ref20}. Interleaving techniques involve generating a sequence of a specific length by combining multiple sequences of predetermined lengths, which may not always be distinct. These methods are widely utilized in sequence construction, particularly for achieving good correlation and developing complementary sequences, among other types of sequences. This subsection introduces and discusses interleaved sequences.

\begin{definition}
    Let \(\mathbf{a}_k = (a_k(0), a_k(1), \ldots, a_k(N-1))\) be a sequence of length \(N\) for \(0 \leq k < M\), and let \(\mathbf{b} = (b(0), b(1), \ldots, b(M-1))\) be a sequence of length \(M\). An \(N \times M\) matrix \(U\) can be constructed by defining its \((k+1)\)-th column as \(b(k) \mathbf{a}^T_k\) for \(0 \leq k < M\). Therefore, we can express \(U\) as:
    \begin{align*}
    U = [b(0) \mathbf{a}^T_0, b(1) \mathbf{a}^T_1, \ldots, b(M-1) \mathbf{a}^T_{M-1}],
   	\end{align*}
    where \(\mathbf{a}^T_k\) denotes the transpose of the sequence \(\mathbf{a}_k\).

    By concatenating the rows of matrix \(U\), we generate an interleaved sequence of length \(MN\). Mathematically, this can be represented as:
    \begin{align}\label{IO}
        \mathbf{u} &= \mathbf{I}(b(0) \mathbf{a}^T_0, b(1) \mathbf{a}^T_1, \ldots, b(M-1) \mathbf{a}^T_{M-1}),
    \end{align}
    where \(\mathbf{I}\) denotes the interleaving operator.
\end{definition}

Let $\mathbf{u}$ be the interleaving sequence defined in (\ref{IO}), and let $\mathbf{v}$ be another interleaving sequence illustrated below:
	\begin{align*}
			\mathbf{v}=\mathbf{I}(c(0)\mathbf{a}^T_0,c(1)\mathbf{a}^T_1,\cdots,c(M-1)\mathbf{a}^T_{M-1}),
	\end{align*}
	where $\mathbf{c}=(c(0),c(1),\cdots,c(M-1))$ is a sequence of length $M$. Let $\tau=M\tau_1+\tau_2, 0\leq \tau_1<N-1,0\leq \tau_2<M$, then the periodic cross-AF of sequence $\mathbf{u}$ and $\mathbf{v}$ at time shift $\tau$ and doppler $v$ is equal to
	\begin{align}\label{ITPAF}
		AF_{\mathbf{u},\mathbf{v}}(\tau,v)
		&=\sum_{n=0}^{M-\tau_2-1}b(n)c^{*}(n+\tau_2)\omega_{NM}^{nv}
		AF_{\mathbf{a}_n,\mathbf{a}_{n+\tau_2}}(\tau_1,v) \nonumber \\		&+\sum_{n=M-\tau_2}^{M-1}b(n)c^{*}(n+\tau_2-M)\omega_{NM}^{nv}
		AF_{\mathbf{a}_n,\mathbf{a}_{n+\tau_2-M}}(\tau_1+1,v).
	\end{align}
	
	The aperiodic AF of sequence $\mathbf{u}$ and $\mathbf{v}$ at time shift $\tau$ and Doppler $v$ is equal to
	\begin{align}\label{ITAAF}
		\hat{AF}_{\mathbf{u},\mathbf{v}}(\tau,v)
		&=\sum_{n=0}^{M-\tau_2-1}b(n)c^{*}(n+\tau_2)\omega_{NM}^{nv}
		\hat{AF}_{\mathbf{a}_n,\mathbf{a}_{n+\tau_2}}(\tau_1,v) \nonumber \\		&+\sum_{n=M-\tau_2}^{M-1}b(n)c^{*}(n+\tau_2-M)\omega_{NM}^{nv}
		\hat{AF}_{\mathbf{a}_n,\mathbf{a}_{n+\tau_2-M}}(\tau_1+1,v).
	\end{align}
	
	\section{Locally Perfect Nonlinear Functions}\label{section-LPNF}
	
In this section, we introduce a novel concept based on perfect nonlinear functions that we discussed earlier. This concept will play a vital role in our development of sequence sets. More specifically, we will present a new class of functions that demonstrate the property of being ``perfectly nonlinear" within a specific local region in both the domain and the co-domain. We call this class of functions \textit{locally perfect nonlinear functions (LPNFs)}.

	%Let $\mathbb{Z}_N$ denote the ring of integers modulo $N$, where $N$ is a positive integer. We define the LPNFs over ring of integers, as follows.

	\begin{definition}\label{GPNFs}
    	Let $N$ and $K$ be two positive integers with $K>N$.  Let $f$ be a function that maps from $\mathbb{Z}_N\) to \(\mathbb{Z}_K$. Define sets $C$ and $D$ as follows: 
    	 
    $$C = \{a \mid -Z_x < a < Z_x\}, \quad D = \{b \mid -Z_y < b < Z_y\},$$  
    
    where $0 < Z_x \leq N$ and $0 < Z_y \leq K$. We define a measure of nonlinearity of $f$ as:
 
    \begin{align*}
			P_f=\max _{0 \neq a \in C} \max _{b \in D}|\{x \in \mathbb{Z}_N: f(x+a)-f(x)=b\}|.
		\end{align*}

    When $P_f = 1$, we say that $f$ is a locally perfect nonlinear function (LPNF) over the set $C$. Additionally, when  $P_f=1,$ for $Z_x=N$ and $Z_y=K$, the function becomes a perfect nonlinear function (PNF).
\end{definition}

%We illustrate this through an explicit example.
	
	\begin{example}
		Let $N=5,K=8,$ $f(x)=\phi(g(x))$ be a function from $\mathbb{Z}_5$ to $\mathbb{Z}_8,$ where $g(x)=x^2$ is the function from $\mathbb{Z}_5$ to $\mathbb{Z}_5$ and $\phi(x)=x$ is the function from $\mathbb{Z}_5$ to $\mathbb{Z}_8.$ By the definition of $f(x),$ we have 
		\begin{equation}\label{ex1}
				\begin{aligned}
					&[x \in \mathbb{Z}_5: f(x+1)-f(x)]=[1,3,0,5,7],  \\
					&[x \in \mathbb{Z}_5: f(x+2)-f(x)]=[4,3,5,4,0], \\
					&[x \in \mathbb{Z}_5: f(x+3)-f(x)]=[4,0,4,5,3], \\
					&[x \in \mathbb{Z}_5: f(x+4)-f(x)]=[1,7,5,0,3]. 
				\end{aligned}
			\end{equation}
			When $\tau=2$ and $\tau=3,$ we have 
			$f(0+2)-f(0)=4,f(3+2)-f(3)=4$ and $f(0+3)-f(0)=4,f( 2+3 )-f(2)=4$. So $f(x+2)-f(x)=4$ and $f( x+3)-f(x)=4$ both have two solutions for $x\in \mathbb{Z}_5.$ By Definition \ref{PNFs}, the function $f$ is not a PNF. It is clear that for $ C=\{a|-5<a<5\}$ and $D=\{b|-4<b<4\}$ by (\ref{ex1}), 	
			\begin{align*}
				P_f=\max _{0 \neq a \in C} \max _{b \in D}|\{x \in \mathbb{Z}_5: f(x+a)-f(x)=b\}|=1.
			\end{align*} 
			Hence $f$ is an LPNF over $\mathbb{Z}_5$. 
	\end{example}

\begin{lemma}\label{function}
    Let $N$ be an odd positive integer and $K > N$ be a positive integer. Let $p$ be the smallest prime factor of  $N$. Define a function $g(x) =a_2 x^2+a_1x$ from $\mathbb{Z}_N$ to $\mathbb{Z}_N$, where $\gcd(a_2, N) = 1$ and $a_1 \in \mathbb{Z}_N$. Define a function $\phi(x) = x$ from $\mathbb{Z}_N $ to $\mathbb{Z}_K$. Then $f(x)$ defined as $f= \phi(g(x))$ maps from $\mathbb{Z}_N$ to $\mathbb{Z}_K$. We have the following cases:

  \begin{itemize}
			\item When $K=N,$ for any $-p<a\neq 0<p $ and $-N<b<N$, $f(x+a)  - f(x)= b$ has at most one solution for $x \in \mathbb{Z}_N$.
			\item When $N< K<2N-1,$ for any $-p<a\neq 0<p $ and $-K+N-1<b<K-N+1$, $f(x+a)-f(x)=b$ has at most one solution for $x \in \mathbb{Z}_N$.
			\item When $K\geq 2N-1,$ for any $-p<a\neq 0<p $ and $-K<b<K$, $f(x+a)-f(x)=b$ has at most one solution for $x \in \mathbb{Z}_N$.
		\end{itemize}
\end{lemma}
		\begin{IEEEproof}
		
		  We will consider the following cases.
		 
Case 1: $K = N$
In this case, we have

\[
f(x+a) - f(x) \equiv \phi(g(x+a)) - \phi(g(x)) \equiv g(x+a) - g(x).
\]

This can be expressed as:

\[
f(x+a) - f(x) = a_2(x+a)^2 + a_1(x+a) - a_2x^2 - a_1x = 2a_2ax + a_2a^2 + a_1a.
\]

Thus, the equation \( f(x+a) - f(x) = b \) simplifies to

\[
2a_2ax = b - a_2a^2 - a_1a. \tag{1}
\]

Given that \( N \) is odd, \( \gcd(a_2, N) = 1 \), and \( -p < a \neq 0 < p \) where \( p \) is the smallest prime factor of \( N \), we recognize that \( 2a_2a \) has a multiplicative inverse in the ring \( \mathbb{Z}_N \). Therefore, equation (1) has a unique solution given by

\[
x = (2a_2a)^{-1}(b - a_2a^2 - a_1a).
\]

Case 2:  $N < K < 2N - 1$
Here, we have

\[
f(x+a) - f(x) = \phi(g(x+a)) - \phi(g(x)) = 
\begin{cases} 
K + g(x+a) - g(x), & \text{if } g(x+a) < g(x), \\ 
g(x+a) - g(x), & \text{if } g(x+a) \geq g(x). 
\end{cases}
\]

Let \( A_{a,1} = \{ x \in \mathbb{Z}_N : g(x+a) < g(x) \} \) and \( A_{a,2} = \{ x \in \mathbb{Z}_N : g(x+a) \geq g(x) \} \). 

For any \( x_1, x_2 \in A_{a,1} \), we find that 

\[
K + g(x_1 + a) - g(x_1) = K + g(x_2 + a) - g(x_2)
\]

if and only if \( x_1 = x_2 \) (as established in Case 1). Similarly, for any \( x_3, x_4 \in A_{a,2} \),

\[
g(x_3 + a) - g(x_3) = g(x_4 + a) - g(x_4)
\]

if and only if \( x_3 = x_4 \).

Next, let \( B_{a,1} = \{ K + g(x + a) - g(x) : x \in A_{a,1} \} \) and \( B_{a,2} = \{ g(x + a) - g(x) : x \in A_{a,2} \} \). It follows that 

\[
|B_{a,1} \cup B_{a,2}| = N.
\] 

Thus, the equation \( f(x+a) - f(x) = b \) has at most one solution for \( x \in \mathbb{Z}_N \) if for any \( -p < a \neq 0 < p \) and \( -K + N - 1 < b < K - N + 1 \), we have

\[
b \not\in B_{a,1} \cap B_{a,2}.
\]

For any \( y_1 \in B_{a,1} \) and \( y_2 \in B_{a,2} \), we can conclude that \( y_1 \geq K - N + 1 \) and \( 0 < y_2 \leq N - 1 \). Thus, \( b \not\in B_{a,1} \cap B_{a,2} \) holds true when \( -K + N - 1 < b < K - N + 1 \).

Case 3:  $K \geq 2N - 1$ 
This case is similar to Case 2. For any \( y_1 \in B_{a,1} \) and \( y_2 \in B_{a,2} \), we have \( y_1 \geq N \) and \( 0 < y_2 \leq N - 1 \). Therefore, for any \( -p < a \neq 0 < p \) and \( -K < b < K \), we can assert that 

\[
b \not\in B_{a,1} \cap B_{a,2}.
\]

This completes the proof.
\end{IEEEproof}

	\section{Constructions and Performance Analysis of LAZ Sequence Set Using LPNFs}\label{Constructions}
	
Our primary focus in this section is on sequence designs, where we will describe our constructions of derived sequence sets. We will evaluate the quality of the generated LAZ sequence sets concerning optimality after determining their parameters, analyzing bounds, and demonstrating that these sets are asymptotically optimal. In particular, we will present LAZ sequence sets that are asymptotically optimal with respect to both the periodic and aperiodic LAZ bounds proposed in  \cite{ye22}. Additionally, we will provide examples to illustrate our results.

	\begin{construction}\label{c1}
		Let $N>2$ be an odd positive integer and $K$ be a positive integer. Let $f:\mathbb{Z}_N \rightarrow \mathbb{Z}_K$ be an LPNF that satisfies $f(x+a)-f(x)=b$ with at most one solution for any $-Z_x<a \neq 0<Z_x,   -Z_y<b<Z_y, $ and $ x \in \mathbb{Z}_N $, where $0<Z_x\leq N,0<Z_y\leq K$. Let $\mathcal{A}$ be an $N\times K$ sequence set defined as follows:
		\begin{equation}
			\mathcal{A}=\begin{bmatrix}
				\mathbf{a}_0\\
				\mathbf{a}_1\\
				\vdots\\
				\mathbf{a}_{(N-1)}
			\end{bmatrix}=\begin{bmatrix}
				a_0(0) & a_0(1) & \hdots & a_0(K-1)\\
				a_1(0) & a_1(1) & \hdots & a_1(K-1)\\
				\vdots & \vdots & \ddots & \vdots\\
				a_{N-1}(0) & a_{N-1}(1) & \hdots & a_{N-1}(K-1)
			\end{bmatrix},
		\end{equation}
		where $a_k(t)=\omega_K^{t f(k)}$. Let $\mathcal{H}$ be another $N\times N$ sequence set defined as
		\begin{equation}
			\mathcal{H}=\begin{bmatrix}
				\mathbf{h}_0\\
				\mathbf{h}_1\\
				\vdots\\
				\mathbf{h}_{(N-1)}
			\end{bmatrix}=\begin{bmatrix}
				h_0(0) & h_0(1) & \hdots & h_0(N-1)\\
				h_1(0) & h_1(1) & \hdots & h_1(N-1)\\
				\vdots & \vdots & \ddots & \vdots\\
				h_{N-1}(0) & h_{N-1}(1) & \hdots & h_{N-1}(N-1)
			\end{bmatrix},
		\end{equation}
		such that
		\begin{equation}
			\begin{split}
				&\left|\sum_{n=0}^{N-1}h_i(n)h^{*}_j(n)\right|\leq 1,\\
				&\left|\sum_{n=0}^{N-1}h_i(n)h^{*}_j(n)\omega_{N}^{nv}\right|<N,
			\end{split}
		\end{equation}
		where $ 0\leq i\neq j<N,0\leq v<N$. Define a sequence set $\mathcal{S}$ of size $N\times NK$ as follows:
		\begin{equation}
			\mathcal{S}=\begin{bmatrix}
				\mathbf{s}_0\\
				\mathbf{s}_1\\
				\vdots\\
				\mathbf{s}_{(N-1)}
			\end{bmatrix}=\begin{bmatrix}
				s_0(0) & s_0(1) & \hdots & s_0(NK-1)\\
				s_1(0) & s_1(1) & \hdots & s_1(NK-1)\\
				\vdots & \vdots & \ddots & \vdots\\
				s_{N-1}(0) & s_{N-1}(1) & \hdots & s_{N-1}(NK-1)
			\end{bmatrix},
		\end{equation}
		where $\mathbf{s}_n=\mathbf{I}(h_n(0)\mathbf{a}^T_0,h_n(1)\mathbf{a}^T_1,\cdots,h_n(N-1)\mathbf{a}^T_{N-1})$, for $0\leq n <N$.
	\end{construction}

	\begin{lemma}
	Each sequence $\mathbf{s}_i$ for $0 \leq i \leq N-1$ in $\mathcal{S}$ is cyclically distinct.
\end{lemma}

\begin{IEEEproof} Let $\mathbf{s}_i$ and $\mathbf{s}_j$ be any two sequences in $\mathcal{S}$, where $0 \leq i \neq j<N$. Assume on the contrary that they are cyclically shifting equivalent, then this is equivalent to the existence of $(\tau,v) \in (-NK,NK) \times (-NK,NK)$ such that $|AF_{\mathbf{s}_i,\mathbf{s}_j}(\tau,v)| = NK$.
	
	Let time shift $\tau=N\tau_1+\tau_2$ and Doppler $v, 0\leq \tau_1<K,0\leq \tau_2<N, 0\leq v<NK,$  the periodic cross AF of $s_i$ and $s_j$ is equal to
	\begin{align}\label{cyd1}
		AF_{\mathbf{s}_i,\mathbf{s}_j}(\tau,v)
		&=\sum_{n=0}^{N-\tau_2-1}h_i(n)h_j^{*}(n+\tau_2)\omega_{NK}^{nv}
		AF_{\mathbf{a}_n,\mathbf{a}_{n+\tau_2}}(\tau_1,v) \nonumber \\		&+\sum_{n=N-\tau_2}^{N-1}h_i(n)h_j^{*}(n+\tau_2-N)\omega_{NK}^{nv}
		AF_{\mathbf{a}_n,\mathbf{a}_{n+\tau_2-N}}(\tau_1+1,v).
	\end{align}
	
	Next, let us analyze \( AF_{\mathbf{a}_n,\mathbf{a}_{n+\tau_2}}(\tau_1,v) \) and \( AF_{\mathbf{a}_n,\mathbf{a}_{n+\tau_2-N}}(\tau_1+1,v) \).
	
	\begin{align}\label{cyd2}
		AF_{\mathbf{a}_n,\mathbf{a}_{n+\tau}}(\tau_1,v)&=\sum_{t=0}^{K-1}a_n(t)a_{n+\tau_2}^{*}(t+\tau_1)\omega_K^{vt} \nonumber \\
		&=\sum_{t=0}^{K-1}\omega_K^{tf(n)}\omega_K^{-(t+\tau_1)f(n+\tau_2)}\omega_K^{vt} \nonumber \\
		&=\omega_{K}^{-\tau_1f(n+\tau_2)}\sum_{t=0}^{K-1}\omega_K^{(f(n)+v-f(n+\tau_2))t} \nonumber \\
		&=\omega_{K}^{-\tau_1f(n+\tau_2)}\delta_K(f(n)+v-f(n+\tau_2)),
	\end{align}
	where the function $\delta_k$ is defined as 
	\[	\delta_K(x)= \begin{cases}K, &  x\equiv 0\mod K, \\ 0, & \text { otherwise.} \end{cases}\]
	
	\begin{align}\label{cyd3}
		AF_{\mathbf{a}_n,\mathbf{a}_{n+\tau_2-N}}(\tau_1+1)&=\sum_{t=0}^{K-1}a_n(t)a_{n+\tau_2-N}^{*}(t+\tau_1+1)\omega_K^{vt}\nonumber \\
		&=\sum_{t=0}^{K-1}\omega_K^{tf(n)}\omega_K^{-(t+\tau_1+1)f(n+\tau_2-N)}\omega_K^{vt}\nonumber \\
		&=\omega_K^{-(\tau_1+1)f(n+\tau_2)}\sum_{t=0}^{K-1}\omega_K^{(f(n)+v-f(n+\tau_2))t} \nonumber \\
		&=\omega_K^{-(\tau_1+1)f(n+\tau_2)}\delta_K(f(n)+v-f(n+\tau_2))
	\end{align}
	
	By (\ref{cyd2}) and (\ref{cyd3}), (\ref{cyd1}) becomes
	
	\begin{align}
		AF_{\mathbf{s}_i,\mathbf{s}_j}(\tau,v)
		&=\sum_{n=0}^{N-\tau_2-1}h_i(n)h_j^{*}(n+\tau_2)\omega_{NK}^{nv}
		\omega_{K}^{\tau_1f(n+\tau_2)}\delta_K(f(n)+v-f(n+\tau_2)) \nonumber \\		&+\sum_{n=N-\tau_2}^{N-1}h_i(n)h_j^{*}(n+\tau_2-N)\omega_{NK}^{nv}
		\omega_K^{-(\tau_1+1)f(n+\tau_2)}\delta_K(f(n)+v-f(n+\tau_2)).
	\end{align}
	
	If $|AF_{\mathbf{s}_i,\mathbf{s}_j}(\tau,v)|=NK,$  then $\delta_K(f(n)+v-f(n+\tau_2))=K$ for all $n\in \mathbb{Z}_N,$ i.e., $K|v,\tau_2=0.$ Let $v=Kv_1,$ then we have 
	\begin{align*}
		AF_{\mathbf{s}_i,\mathbf{s}_j}(\tau,v)=	AF_{\mathbf{s}_i,\mathbf{s}_j}(\tau,Kv_1)=K\sum_{n=0}^{N-1}h_i(n)h_j^{*}(n)\omega_{N}^{nv_1} .
	\end{align*}
	%主要的目的就是让$|\sum_{n=0}^{N-1}O_i(n)O_j^{*}(n)\omega_{N}^{nk}|<N.$
	
	If $|AF_{\mathbf{s}_i,\mathbf{s}_j}(\tau,v)|=NK$, then $\left|\sum_{n=0}^{N-1}h_i(n)h_j^{*}(n)\omega_{N}^{nv_1}\right|=N$. This contradicts the inequality 
	\\$\left|\sum_{n=0}^{N-1}h_i(n)h_j^{*}(n)\omega_{N}^{nv_1}\right|<N$
	for $0\leq v_1<N$. Therefore, we conclude that all the sequences in $\mathcal{S}$ are cyclically distinct.  The proof is completed.
\end{IEEEproof}
	
	We have the following theorem regarding the sequence set $\mathcal{S}$  which is obtained in Construction \ref{c1}.
	
	\begin{theorem}\label{ATh1}
	The sequence set $\mathcal{S}$, constructed using Construction \ref{c1}, is a polyphase periodic LAZ sequence set with parameters $(N, NK, \Pi, K)$, where $\Pi = (-Z_x, Z_x) \times (-Z_y, Z_y)$.
	\end{theorem}
	
	\begin{IEEEproof}
		Let $\mathbf{s}_i$ and $\mathbf{s}_j$ be any two sequences in $\mathcal{S}$, where $0 \leq i, j < N$. For a delay shift $\tau$ and a Doppler shift $v$, where $0 \leq \tau < Z_x < N$ and $0 \leq v < NK$, we have the periodic cross autocorrelation function of $\mathbf{s}_i$ and $\mathbf{s}_j$ as follows:
		\begin{align}\label{6}
			AF_{\mathbf{s}_i,\mathbf{s}_j}(\tau,v)
			&=\sum_{n=0}^{N-\tau-1}h_i(n)h_j^{*}(n+\tau)\omega_{NK}^{nv}
			AF_{\mathbf{a}_n,\mathbf{a}_{n+\tau}}(0,v) \nonumber \\		&+\sum_{n=N-\tau}^{N-1}h_i(n)h_j^{*}(n+\tau-N)\omega_{NK}^{nv}
			AF_{\mathbf{a}_n,\mathbf{a}_{n+\tau-N}}(1,v).
		\end{align}
		
		By  (\ref{cyd2}) and  (\ref{cyd3}), we have 
			\begin{align*}
				&AF_{\mathbf{a}_n,\mathbf{a}_{n+\tau}}(0,v)=\delta_K(f(n)+v-f(n+\tau)),\\
				&AF_{\mathbf{a}_n,\mathbf{a}_{n+\tau-N}}(1,v)=\omega_K^{-f(n+\tau)}\delta_K(f(n)+v-f(n+\tau)).
		\end{align*}

		Thus,  (\ref{6}) becomes
			\begin{align}
				AF_{\mathbf{s}_i,\mathbf{s}_j}(\tau,v)
				&=\sum_{n=0}^{N-\tau-1}h_i(n)h_j^{*}(n+\tau)\omega_{NK}^{nv}
				\delta_K(f(n)+v-f(n+\tau)) \nonumber \\		&+\sum_{n=N-\tau}^{N-1}h_i(n)h_j^{*}(n+\tau-N)\omega_{NK}^{nv}
				\omega_K^{-f(n+\tau)}\delta_K(f(n)+v-f(n+\tau)).
		\end{align}
		
	We shall distinguish the following two cases.
		
		Case 1: For $i=j$, we have
		\begin{align}
    AF_{\mathbf{s}_i,\mathbf{s}_i}(\tau,v)
    &= \sum_{n=0}^{N-\tau-1} h_i(n) h_i^{*}(n+\tau) \omega_{NK}^{nv} \delta_K(f(n)+v-f(n+\tau)) \nonumber \\ 
    & + \sum_{n=N-\tau}^{N-1} h_i(n) h_i^{*}(n+\tau) \omega_{NK}^{nv} \omega_K^{-f(n+\tau)} \delta_K(f(n)+v-f(n+\tau)).
\end{align}

When \(\tau=0\) and \(v=0\), we find that
\begin{align*}
    AF_{\mathbf{s}_i,\mathbf{s}_i}(0,0) = \sum_{n=0}^{N-1} h_i(n) h_i^{*}(n) AF_{\mathbf{a}_n,\mathbf{a}_{n}}(0,0) = NK.
\end{align*}

For the case when \(\tau=0\) and \(0<|v|<NK\), we have 
\begin{align*}
    AF_{\mathbf{s}_i,\mathbf{s}_i}(0,v) = \sum_{n=0}^{N-1} h_i(n) h_i^{*}(n) \omega_{NK}^{nv} \delta_K(f(n)+v-f(n)) = \sum_{n=0}^{N-1} \omega_{NK}^{nv} \delta_K(v) = 0.
\end{align*}

If \( -Z_x < a \neq 0 < Z_x\) and \(-Z_y < b < Z_y\), then the equation \(f(x+a) - f(x) = b\) has at most one solution for \(x \in \mathbb{Z}_N\). Thus, we can conclude that
\begin{align*}
    |AF_{\mathbf{s}_i}(\tau,v)| \leq K.
\end{align*}
		
		Case 2: For $i\neq j,$ we have
			\begin{align}
				AF_{\mathbf{s}_i,\mathbf{s}_j}(\tau,v)
				&=\sum_{n=0}^{N-\tau-1}h_i(n)h_j^{*}(n+\tau)\omega_{NK}^{nv}
				\delta_K(f(n)+v-f(n+\tau)) \nonumber \\		&+\sum_{n=N-\tau}^{N-1}h_i(n)h_j^{*}(n+\tau-N)\omega_{NK}^{nv}
				\omega_K^{-f(n+\tau)}\delta_K(f(n)+v-f(n+\tau)).
		\end{align}
		
When \(\tau = 0\) and \(v = 0\), we have:

\[|AF_{\mathbf{s}_i,\mathbf{s}_j}(\tau,v)|=\left|\sum_{n=0}^{N-1}h_i(n)h_j^{*}(n)AF_{\mathbf{a}_n,\mathbf{a}_{n}}(0,0)\right|=\left|K\sum_{n=0}^{N-1}h_i(n)h_j^{*}(n)\right|\leq K.
\]

When \(\tau = 0\) and \(0 < |v| < Z_y\), we obtain:

\[
AF_{\mathbf{s}_i,\mathbf{s}_j}(\tau,v) = \sum_{n=0}^{N-1} h_i(n) h_j^*(n) \omega_{NK}^{nv} \delta_K(f(n) + v - f(n)) = \sum_{n=0}^{N-1} h_i(n) h_j^*(n) \omega_{NK}^{nv} \delta_K(v) = 0.
\]

When \(-Z_x < a \neq 0 < Z_x\) and \(-Z_y < b < Z_y\), the equation \(f(x + a) - f(x) = b\) has at most one solution for \(x \in \mathbb{Z}_N\). Consequently, we have:

\[
|AF_{\mathbf{s}_i,\mathbf{s}_j}(\tau,v)| \leq K.
\]

Therefore, the sequence set \(\mathcal{S}\) is a polyphase periodic \((N, NK, \Pi, K)\)-LAZ sequence set with \(\Pi = (-Z_x, Z_x) \times (-Z_y, Z_y)\), which completes the proof.
	\end{IEEEproof}

	\begin{theorem} \label{T2}
	The sequence set $\mathcal{S}$ created in Construction \ref{c1} is an aperiodic LAZ sequence set with parameters $(N, NK, \Pi, K + Z_x - 1)$, where $\Pi = (-Z_x, Z_x) \times (-Z_y, Z_y)$.
	\end{theorem}
	
	\begin{IEEEproof}
		Let $\mathbf{s}_i$ and $\mathbf{s}_j$ be two arbitrary sequences in $\mathcal{S}$. For time-shift $\tau$ and Doppler shift $v, 0\leq \tau<Z_x<N, 0\leq v<NK$, we have the aperiodic cross-AF of $s_i$ and $s_j$, as follows:
		\begin{align}\label{9}
			\hat{AF}_{\mathbf{s}_i,\mathbf{s}_j}(\tau,v)
			&=\sum_{n=0}^{N-\tau-1}h_i(n)h_j^{*}(n+\tau)\omega_{NK}^{nv}
			\hat{AF}_{\mathbf{a}_n,\mathbf{a}_{n+\tau_2}}(0,v) \nonumber \\		&+\sum_{n=N-\tau}^{N-1}h_i(n)h_j^{*}(n+\tau-N)\omega_{NK}^{nv}
			\hat{AF}_{\mathbf{a}_n,\mathbf{a}_{n+\tau_2-N}}(1,v).
		\end{align}
		Next, let us analyze $\hat{AF}_{\mathbf{a}_n,\mathbf{a}_{n+\tau_2}}(0,v)$ and $\hat{AF}_{\mathbf{a}_n,\mathbf{a}_{n+\tau_2-N}}(1,v).$
		
		\begin{align}\label{10}
				\hat{AF}_{\mathbf{a}_n,\mathbf{a}_{n+\tau_2}}(0,v)=AF_{\mathbf{a}_n,\mathbf{a}_{n+\tau_2}}(0,v)=\delta_K(f(n)+v-f(n+\tau)).
		\end{align}
		
		\begin{align}\label{11}
				\hat{AF}_{\mathbf{a}_n,\mathbf{a}_{n+\tau_2-N}}(1,v)&=\sum_{t=0}^{K-2}a_n(t)a_{n+\tau-N}^{*}(t+1)\omega_K^{vt} \nonumber \\
				&=\sum_{t=0}^{K-2}\omega_K^{tf(n)}\omega_K^{-(t+1)f(n+\tau-N)}\omega_K^{vt} \nonumber \\
				&=\omega_K^{-f(n+\tau)}\sum_{t=0}^{K-1}\omega_K^{(f(n)+v-f(n+\tau))t}-\omega_{K}^{-f(n)-v} \nonumber \\
				&=\omega_K^{-f(n+\tau)}\delta_K(f(n)+v-f(n+\tau))-\omega_{K}^{-f(n)-v}
			\end{align}
			By  (\ref{10}) and  (\ref{11}),  (\ref{9}) becomes
			\begin{align}
				\hat{AF}_{\mathbf{s}_i,\mathbf{s}_j}(\tau,v)
				&=\sum_{n=0}^{N-\tau-1}h_i(n)h_j^{*}(n+\tau)\omega_{NK}^{nv}
				\delta_K(f(n)+v-f(n+\tau)) \nonumber \\		&+\sum_{n=N-\tau}^{N-1}h_i(n)h_j^{*}(n+\tau)\omega_{NK}^{nv}
				\omega_K^{-f(n+\tau)}\delta_K(f(n)+v-f(n+\tau))\nonumber \\
				&-\sum_{n=N-\tau}^{N-1}h_i(n)h_j^{*}(n+\tau)\omega_{NK}^{nv}\omega_{K}^{-f(n)-v}.
		\end{align}
		
		Consequently,
		\begin{align}
			|\hat{AF}_{\mathbf{s}_i,\mathbf{s}_j}(\tau,v)|&\leq |AF_{\mathbf{s}_i,\mathbf{s}_j}(\tau,v)|+\left|\sum_{n=N-\tau}^{N-1}h_i(n)h_j^{*}(n+\tau)\omega_{NK}^{nv}\omega_{K}^{-f(n)-v}\right|\nonumber \\
			& \leq |AF_{\mathbf{s}_i,\mathbf{s}_j}(\tau,v)|+\tau \nonumber \\
			&\leq |AF_{\mathbf{s}_i,\mathbf{s}_j}(\tau,v)|+Z_x-1
		\end{align}
		
		According to  Theorem \ref{ATh1}, we have $|AF_{\mathbf{s}_i,\mathbf{s}_j}(\tau,v)|\leq K$ in region $(-Z_x,Z_x)\times (-Z_y,Z_y)$ and $|AF_{\mathbf{s}_i}(\tau,v)|\leq K$ in region $(-Z_x,Z_x)\times (-Z_y,Z_y) $ except at the origin of the DD domain. It can be derived that $|\hat{AF}_{\mathbf{s}_i,\mathbf{s}_j}(\tau,v)|\leq K+Z_x-1$ in region $(-Z_x,Z_x)\times (-Z_y,Z_y)$.%and  $|\hat{AF}_{\mathbf{s}_i}(\tau,v)|\leq K+Z_x-1$ in region $(-Z_x,Z_x)\times (-Z_y,Z_y)$ except at the origin of the DD domain.
		
		Therefore, the set \( \mathcal{S} \) is a polyphase aperiodic \((N, NK,\Pi, K+Z_x-1)\)-LAZ sequence set with \(\Pi = (-Z_x,Z_x)\times (-Z_y,Z_y)\), which completes the proof.
	\end{IEEEproof}

Using the function given in Lemma \ref{function} in Theorem \ref{ATh1}, we have the following result, which can be proved similarly as for establishing Theorem \ref{ATh1}. Hence, we 
omit the proof.

	\begin{corollary} \label{Th3}
		The sequence set $\mathcal{S}$ designed in Construction \ref{c1} using the LPNF given in Lemma \ref{function}, is a periodic LAZ sequence set with the following parameters:
		%	By using the function in Lemma \ref{function},  the periodic LAZ sequence set $\mathcal{S}$ constructed using Construction \ref{c1} with the following parameters:
		\begin{itemize}
			\item  When $K=N,$ then $\mathcal{S}$ is an $(N,NK,\Pi,K)$ with $\Pi=(-p,p)\times (-N,N);$
			\item  When $N<K<2N-1,$ then $\mathcal{S}$ is an $(N,NK,\Pi,K)$ with $\Pi=(-p,p)\times (-K+N-1,K-N+1);$ 
			\item  When $K\geq 2N-1,$ then $\mathcal{S}$ is an $(N,NK,\Pi,K)$ with $\Pi=(-p,p)\times (-K,K).$
		\end{itemize}
	\end{corollary}

	Using the function given in Lemma \ref{function} in Theorem \ref{T2}, we have the following corollary.
	
	\begin{corollary}\label{Th4}
		The sequence set $\mathcal{S}$ designed in Construction \ref{c1} using the LPNF given in Lemma \ref{function}, is an aperiodic LAZ sequence set with the following parameters:
		\begin{itemize}
			\item  When $K=N,$ then $\mathcal{S}$ is an $(N,NK,\Pi,K+p-1)$ with $\Pi=(-p,p)\times (-N,N);$
			\item  When $N<K<2N-1,$ then $\mathcal{S}$ is an $(N,NK,\Pi,K+p-1)$ with $\Pi=(-p,p)\times (-K+N-1,K-N+1);$ 
			\item  When $K\geq 2N-1,$ then $\mathcal{S}$ is an $(N,NK,\Pi,K+p-1)$ with $\Pi=(-p,p)\times (-K,K).$
		\end{itemize}
	\end{corollary}

	\begin{remark}
	Recall the two constraints on the sequence set $\mathcal{H}$ in Construction \ref{c1}. The sequence set $\mathcal{H}$ is defined as an $N \times N$ matrix:
		\begin{equation}
			\mathcal{H}=\begin{bmatrix}
				\mathbf{h}_0\\
				\mathbf{h}_1\\
				\vdots\\
				\mathbf{h}_{(N-1)}
			\end{bmatrix}=\begin{bmatrix}
				h_0(0) & h_0(1) & \hdots & h_0(N-1)\\
				h_1(0) & h_1(1) & \hdots & h_1(N-1)\\
				\vdots & \vdots & \ddots & \vdots\\
				h_{N-1}(0) & h_{N-1}(1) & \hdots & h_{N-1}(N-1)
			\end{bmatrix},
		\end{equation}
		such that
		\begin{equation}\label{eq28}
			\begin{split}
				&\left|\sum_{n=0}^{N-1}h_i(n)h^{*}_j(n)\right|\leq 1,\\
				&\left|\sum_{n=0}^{N-1}h_i(n)h^{*}_j(n)\omega_{N}^{nv}\right|<N,
			\end{split}
		\end{equation}
		where $0\leq i\neq j<N,0\leq v<N$. We can construct the matrix $\mathcal{H}$ that meets the specified conditions using DFT matrices, $m$-sequences, Legendre sequences, and Bjorck sequences in the following ways:
		\begin{itemize}
		\item From a DFT matrix of size $N+1$, we can select any submatrix of order $N$.
\item An $m$-sequence of length $N$ can be represented as the first row of the matrix $\mathcal{H}$, with each subsequent row being a left circular shift of the first row. Similarly, we can use either a Legendre sequence or a Bjorck sequence of length $N$ as the first row.
		\end{itemize}

		It can be shown that the conditions specified in \eqref{eq28} are met if we can construct the set $\mathcal{H}$ as described above.
	\end{remark}

\begin{remark}
Let  $N = p -1$, $K = p$, and define the function $f(x) = \alpha^x$. The matrix $ \mathcal{H}$  is obtained by selecting a submatrix of order $p - 1$.  from a Discrete Fourier Transform (DFT) matrix of order  $p$. The $ i$-th row of matrix $\mathcal{H}$ is given by  $\mathbf{h}_i = (h_i(0), h_i(1), \ldots, h_i(p - 2))$, where $h_i(k) = \omega_{p}^{ik}$ for $ 0 \leq i, k < N $, and  $\alpha $ is a primitive element of  $\mathbb{Z}_p$. The periodic LAZ sequence set with parameters $(p - 1, p(p - 1), \Pi, p)$ , where $ \Pi = (-p + 1, p - 1) \times (-p, p)$, proposed in \cite{tian2024asymptotically}, can be obtained using Construction \ref{c1}. According to Theorem \ref{T2}, we can also derive an aperiodic LAZ sequence set with parameters $(p - 1, p(p - 1), \Pi, 2p - 2)$, where $\Pi = (-p + 1, p - 1) \times (-p, p)$.
\end{remark}
	
\subsection{Analysis of the Optimality of the Constructed Sequence Sets}

In this subsection, we examine the optimality of the proposed periodic LAZ sequences set \( \mathcal{S} \) mentioned in Corollary \ref{Th3} and Corollary \ref{Th4}, respectively. This analysis is grounded in the bound provided in Lemma \ref{PLAZ}. We shall demonstrate our achievements through the results outlined in Theorem \ref{Th3} and Theorem \ref{Th4}.

	\begin{theorem} Let $\mathcal{S}$ be the periodic LAZ sequence set constructed using Corollary \ref{Th3}. Then
		%	Let $N$ be an odd integer, $K$ be an integer,  $\gamma$ be a real number, and $p$ be the smallest prime factor of $N$. Also let $ \lim\limits_{p \rightarrow \infty}\sqrt{\frac{K}{N}}=\gamma.$ Then
	
	\begin{enumerate}
			\item[i)] $\mathcal{S}$ is asymptotically optimal with respect to the bound in Lemma \ref{PLAZ} if $K=N$.
			\item[ii)] $\mathcal{S}$ is  asymptotically optimal with respect to the bound in Lemma \ref{PLAZ}, if $N<K<2N-1$, and $(K-N+1)p\geq Np^{\frac{1}{n}},$ where $n$ is a finite positive integer.
			\item[iii)] When $K\geq 2N-1$, the optimality factor $\rho_{\text{LAZ}}$ is $\gamma$ times the bound in Lemma \ref{PLAZ}, where $ \lim\limits_{p \rightarrow \infty}\sqrt{\frac{K}{N}}=\gamma\geq \sqrt{2}$.
		\end{enumerate}
	\end{theorem}
	
	\begin{IEEEproof}
	
We shall examine different cases based on the value of K as outlined in Theorem \ref{Th3}.

Case 1: If \( K = N \), based on Theorem \ref{Th3}, the set \( \mathcal{S} \) is a periodic \((N, N^2, \Pi, N)\)-LAZ sequence set with \(\Pi = (-p, p) \times (-N, N)\), where \( p \) is the smallest prime factor of \( N \). The optimality factor \( \rho_{\mathrm{LAZ}} \) is given by:

\[
\rho_{\mathrm{LAZ}} = \frac{N}{\frac{N^2}{\sqrt{N}} \sqrt{\frac{N^2p/N^2 - 1}{Np - 1}}} = \sqrt{\frac{Np - 1}{Np - N}} = \sqrt{1 + \frac{N - 1}{Np - N}}.
\]

We find that 

\[
\lim_{p \to \infty} \rho_{\mathrm{LAZ}} = 1.
\]

Thus, the proposed periodic LAZ sequence set \( \mathcal{S} \) is asymptotically optimal.

Case 2: If \( N < K < 2N - 1 \), based on Theorem \ref{Th3}, the set \( \mathcal{S} \) is a periodic \((N, NK, \Pi, K)\)-LAZ sequence set with \(\Pi = (-p, p) \times (-K + N - 1, K - N + 1)\), where \( p \) is the smallest prime factor of \( N \). The optimality factor \( \rho_{\mathrm{LAZ}} \) is given by:

\[
\rho_{\mathrm{LAZ}} = \frac{K}{\frac{NK}{\sqrt{K - N + 1}} \sqrt{\frac{N(K - N + 1)p/NK - 1}{Np - 1}}} = \sqrt{\frac{(K - N + 1)pK}{(K - N + 1)pN - KN} - \frac{(K - N + 1)K}{(K - N + 1)pN^2 - KN^2}}.
\]

We have 

\[
\lim_{p \to \infty} \rho_{\mathrm{LAZ}} = \lim_{p \to \infty} \sqrt{\frac{(K - N + 1)pK}{(K - N + 1)pN - KN}} = \gamma.
\]

Therefore, the proposed periodic LAZ sequence set \( \mathcal{S} \) is asymptotically optimal when \( \gamma = 1 \).

Case 3:  If \( K \geq 2N - 1 \), based on Theorem \ref{Th3}, the set \( \mathcal{S} \) is a periodic \((N, NK, \Pi, K)\)-LAZ sequence set with \(\Pi = (-p, p) \times (-K, K)\), where \( p \) is the smallest prime factor of \( N \). The optimality factor \( \rho_{\mathrm{LAZ}} \) is given by:

\[
\rho_{\mathrm{LAZ}} = \frac{K}{\frac{NK}{\sqrt{K}} \sqrt{\frac{NKp/NK - 1}{Np - 1}}} = \sqrt{\frac{K}{N}} \sqrt{\frac{Np - 1}{Np - N}} = \sqrt{\frac{K}{N}} \sqrt{1 + \frac{N - 1}{Np - N}}.
\]

Hence, 

\[
\lim_{p \to \infty} \rho_{\mathrm{LAZ}} = \lim_{p \to \infty} \sqrt{\frac{K}{N}} = \gamma.
\]

\end{IEEEproof}

	\begin{theorem} Let $\mathcal{S}$ be the aperiodic LAZ sequence set constructed using Corollary \ref{Th4}. Then
		%Let $ \lim\limits_{p \rightarrow \infty}\sqrt{\frac{K}{N}}=\gamma$ and $N$ is not prime.
		\begin{enumerate}
			\item[i)]  the sequence set $\mathcal{S}$  is asymptotically optimal based on the bound given in Lemma \ref{PLAZ}, when $K=N$ and $N$ is not prime.
			\item[ii)]   the sequence set $\mathcal{S}$  is asymptotically optimal based on the bound given in Lemma \ref{PLAZ}, when $N<K<2N-1$, $(K-N+1)p\geq Np^{\frac{1}{n}},$ and $N$ is not prime, where $n$ is a finite positive integer.
			\item[iii)] When $K\geq 2N-1$ and $N$ is not prime, the optimality factor $\hat{\rho}_{\text{LAZ}}$ is $\gamma$ times the bound in Lemma \ref{PLAZ}, where $ \lim\limits_{p \rightarrow \infty}\sqrt{\frac{K}{N}}=\gamma\geq \sqrt{2}$.
		\end{enumerate}

	\end{theorem}

\begin{IEEEproof}
	We shall examine different cases based on the value of K  according to Theorem \ref{Th4}.

    Case 1: If \( K=N \), based on Theorem \ref{Th4}, the set \(\mathcal{S}\) is an aperiodic \((N,N^2,\Pi,N+p-1)\)-LAZ sequence set with \(\Pi=(-p,p)\times(-N,N)\), where \( p \) is the smallest prime factor of \( N \). The optimality factor in this case is given by
    \begin{align*}
        \hat{\rho}_{\mathrm{LAZ}} &=\frac{N+p-1}{\frac{N^2}{\sqrt{N}} \sqrt{\frac{N^2p - N^2 - p + 1}{\left(N^2 + p - 1\right)\left(Np - 1\right)}}} = \frac{N+p-1}{N\sqrt{N}} \sqrt{\frac{(Np-1)}{p-1} + \frac{p(Np-1)}{(N^2-1)(p-1)}}.
    \end{align*}
    If \( N \) is not prime, then \( p \leq \sqrt{N} \) since \( p \) is the smallest prime factor of \( N \). We have
    \begin{align*}
        \lim_{p \rightarrow \infty} \hat{\rho}_{\mathrm{LAZ}} &= \lim_{p \rightarrow \infty} \frac{N+p-1}{N\sqrt{N}} \sqrt{\frac{(Np-1)}{p-1} + \frac{p(Np-1)}{(N^2-1)(p-1)}} \\
        &= \lim_{p \rightarrow \infty} \frac{1}{\sqrt{N}} \sqrt{\frac{(Np-1)}{p-1}} \\
        &= 1.
    \end{align*}
    Hence, the proposed aperiodic LAZ sequence set \(\mathcal{S}\) is asymptotically optimal.
    
    Case 2: If \( N < K < 2N - 1 \), based on Theorem \ref{Th4}, the set \(\mathcal{S}\) is an aperiodic \((N,NK,\Pi,K+p-1)\)-LAZ sequence set with \(\Pi=(-p,p)\times(-K + N - 1,K - N + 1)\), where \( p \) is the smallest prime factor of \( N \). The optimality factor, in this case,  is given by
    
	\begin{align*}
			\hat{\rho}_{\mathrm{LAZ}}&=\frac{K+p-1}{\frac{NK}{\sqrt{K-N+1}} \sqrt{\frac{N(K-N+1)p -NK-p+1}{\left(NK+p-1\right)\left(Np-1\right)}}}\\&=\frac{(K+p-1)\sqrt{K-N+1}}{N\sqrt{K}}\sqrt{\frac{(NK+p-1)(Np-1)}{N(K-N+1)p-NK-p+1}}.
		\end{align*}
		
 	If $N$ is not prime,  We have $\lim\limits_{p \rightarrow \infty} \hat{\rho}_{\mathrm{LAZ}} = \lim\limits_{p \rightarrow \infty}\sqrt{\frac{K}{N}}=\gamma$. Therefore, the proposed aperiodic LAZ sequence set $\mathcal{S}$ is asymptotically optimal when $\gamma=1$.
    
    Case 3: If \( K \geq 2N - 1 \), based on Theorem \ref{Th4}, the set \(\mathcal{S}\) is an aperiodic \((N,NK,\Pi,K+p-1)\)-LAZ sequence set with \(\Pi=(-p,p)\times(-K,K)\), where \( p \) is the smallest prime factor of \( N \). The optimality factor, in this case, is given by
		
    \begin{align*}
			\hat{\rho}_{\mathrm{LAZ}}=\frac{K+p-1}{\frac{NK}{\sqrt{K}} \sqrt{\frac{NKp -NK-p+1}{\left(NK+p-1\right)\left(Np-1\right)}}}=\frac{K+p-1}{N\sqrt{K}}\sqrt{\frac{(Np-1)}{p-1}+\frac{p(Np-1)}{(NK-1)(p-1)}}.
		\end{align*}
    
Proof similar to case 1, if $N$ is not prime,  We have $\lim\limits_{p \rightarrow \infty} \hat{\rho}_{\mathrm{LAZ}} = \lim\limits_{p \rightarrow \infty}\sqrt{\frac{K}{N}}=\gamma$.
	\end{IEEEproof}
	
%	\subsection{Examples}
%In this subsection, we present some examples to illustrate the proposed constructions. The figures also present the numerical results.

	\begin{example}\label{ex2}
	Let \( K = N = 35 \) and consider the function \( f(x) = x^2 \). The sequence set \( \mathcal{H} \) is selected from the submatrix of order \( N \) of the DFT matrix of order \( N + 1 \). The sequence set \( \mathcal{S} \) is constructed using Construction \ref{c1}. It can be shown that \( \mathcal{S} \) forms a periodic \( (35, 1225, \Pi, 35) \)-LAZ sequence set, where \( \Pi = (-5, 5) \times (-35, 35) \). The optimality factor in this case is \( \rho_{\text{LAZ}} = 1.093344 \). A glimpse of the periodic auto-AF and cross-AF of the sequences in \( \mathcal{S} \) can be observed in Fig. \ref{f1}.
		
		\begin{figure}
			\centering 
			\includegraphics[width=\textwidth]{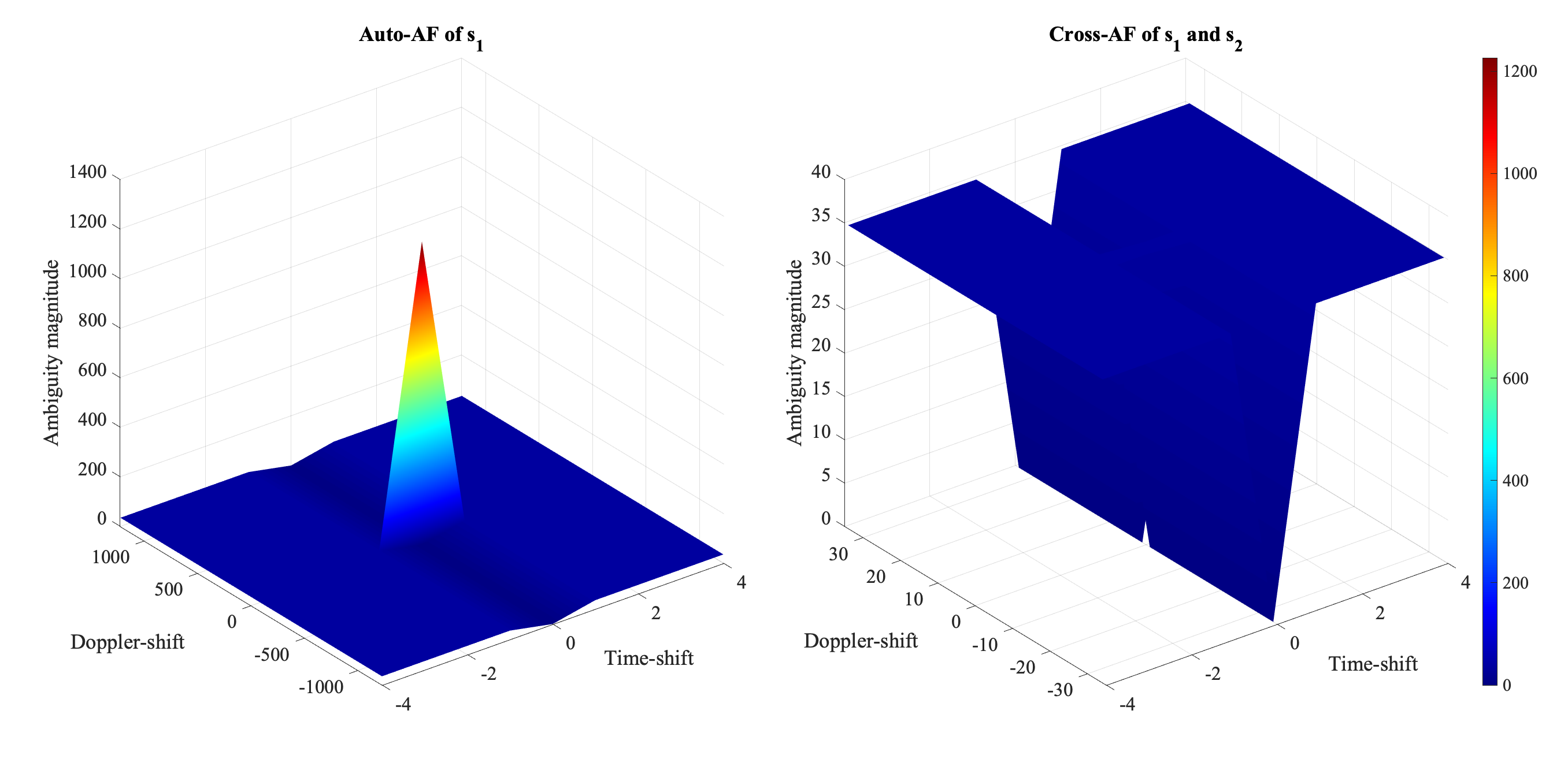}
			\caption{A glimpse of the periodic auto-AF and cross-AF of the sequence set $\mathcal{S}$ in Example \ref{ex2}.}\label{f1}
		\end{figure}

The set \(\mathcal{S}\) can be demonstrated to be an aperiodic \((35, 1225, \Pi, 39)\)-LAZ sequence set, where \(\Pi = (-5, 5) \times (-35, 35)\). In this case, the optimality factor is represented as \(\hat{\rho}_{\text{LAZ}} = 1.244779\). A glimpse of the aperiodic auto-AF and the cross-AF of the sequences in \(\mathcal{S}\) can be seen in Fig. \ref{f2}.	\begin{figure}
			\centering 
			\includegraphics[width=\textwidth]{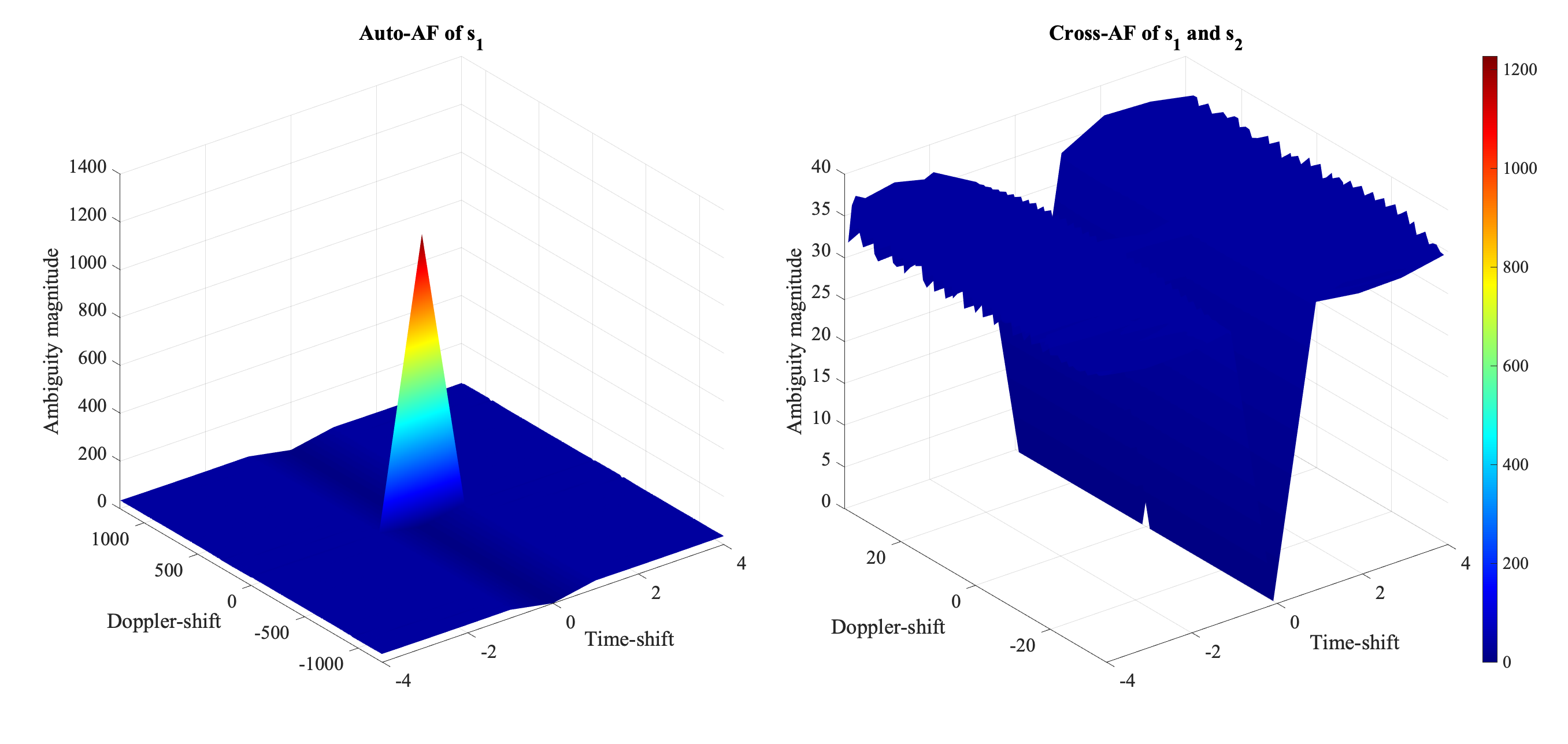}
			\caption{A glimpse of the aperiodic auto-AF and cross-AF of the sequence set $\mathcal{S}$ in Example \ref{ex2}.}\label{f2}
		\end{figure}
	\end{example}

	\begin{example}\label{ex3}
	
	Let \( K = N = 7 \) and consider the function \( f(x) = x^2 \). The sequence set \( \mathcal{H} \) is constructed using all the cyclic shifts of the Legendre sequence. In contrast, the sequence set \( \mathcal{S} \) is created using Construction \ref{c1}. It can be demonstrated that \( \mathcal{S} \) forms a periodic \( (7, 49, \Pi, 7) \)-LAZ sequence set, where \( \Pi = (-7, 7) \times (-49, 49) \). The optimality factor in this case is \( \rho_{\text{LAZ}} = 1.060660 \). A glimpse of the periodic auto-AF and cross-AF of the sequences in \( \mathcal{S} \) can be seen in Fig. \ref{f3}. 
	
\begin{figure}
    \centering 
    \includegraphics[width=\textwidth]{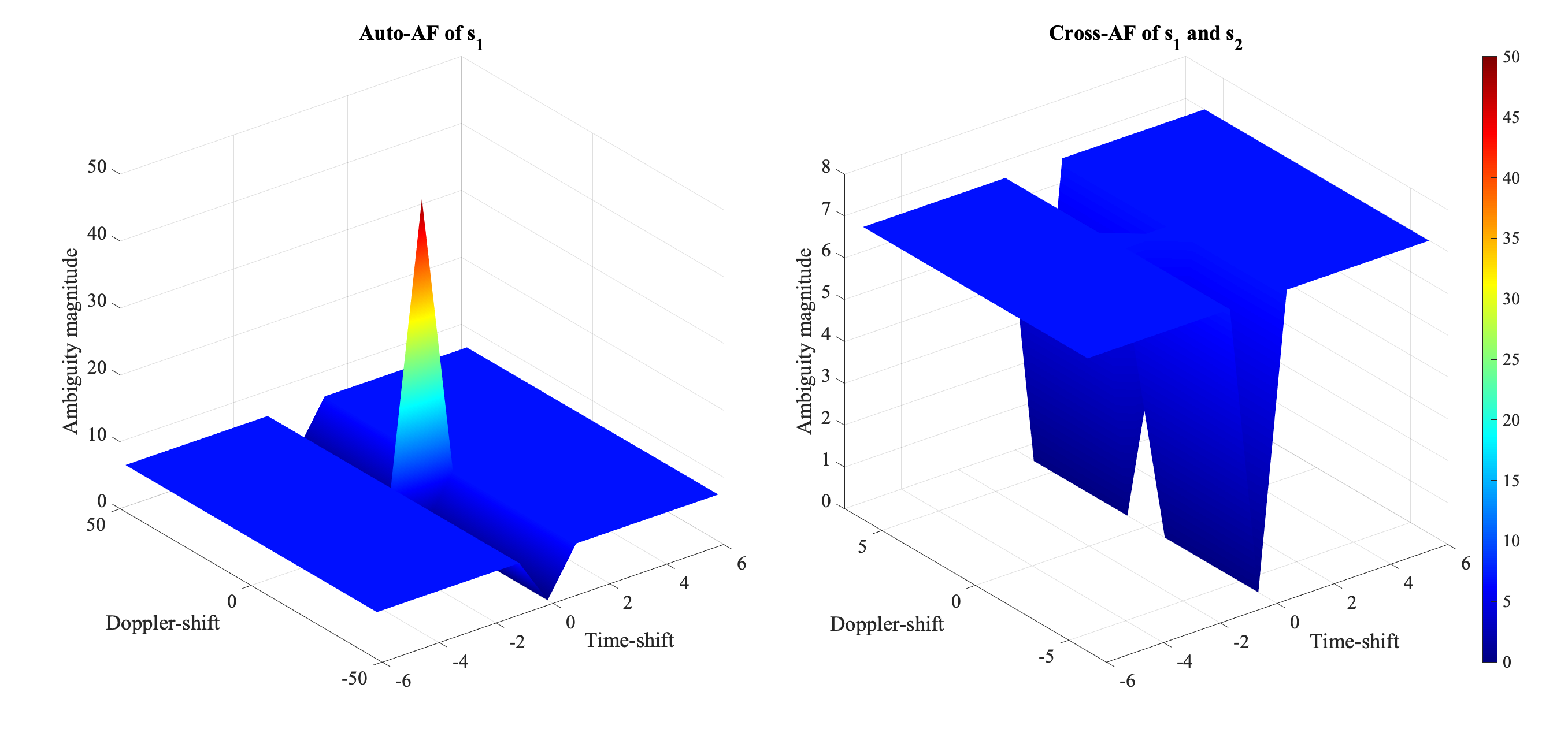}
    \caption{A glimpse of the periodic auto-AF and cross-AF of the sequence set \( \mathcal{S} \) in Example \ref{ex3}.}
    \label{f3}
\end{figure}

Additionally, it can be shown that \( \mathcal{S} \) also constitutes an aperiodic \( (7, 49, \Pi, 13) \)-LAZ sequence set, where \( \Pi = (-7, 7) \times (-49, 49) \). The optimality factor in this scenario is \( \hat{\rho}_{\text{LAZ}} = 2.125211 \). A glimpse of the aperiodic auto-AF and cross-AF of the sequences in \( \mathcal{S} \) can be viewed in Fig. \ref{f4}. 
		
		\begin{figure}
			\centering 
			\includegraphics[width=\textwidth]{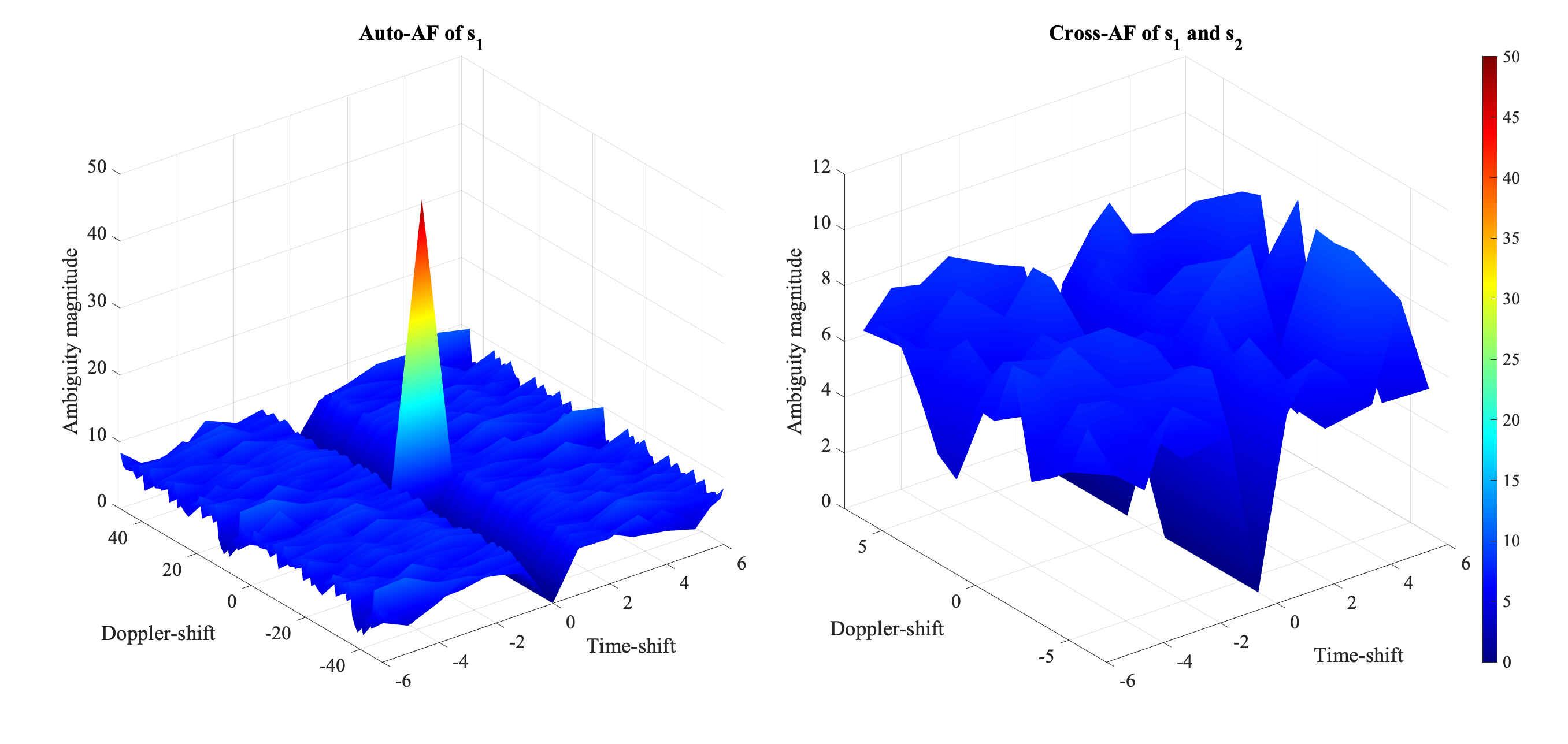}
			\caption{A glimpse of the aperiodic auto-AF and cross-AF of the sequence set $\mathcal{S}$ in Example \ref{ex3}.}\label{f4}
		\end{figure}
		
	\end{example}

	\subsection{Parameters of the Derived Periodic and Aperiodic LAZ Sequence Sets}
	In Tables \ref{table1} and \ref{table2}, we present several parameters of the proposed periodic LAZ sequence sets, along with their corresponding optimality factor, $\rho_{\text{LAZ}}$, for the cases when \( K = N \) and \( N < K < 2N - 1 \), respectively. Additionally, we provide parameters of the proposed aperiodic LAZ sequence sets for the cases when \( K = N \) and \( N < K < 2N - 1 \), along with their corresponding optimality factor, \(\hat{\rho}_{\text{LAZ}}\), in Tables \ref{2} and \ref{3}, respectively.

	\begin{remark}
We emphasize that the parameters of the proposed construction do not meet the conditions outlined in \cite{meng2024new}. Therefore, we have adopted the aperiodic Ye-Zhou-Liu-Fan-Lei-Tang bound to evaluate the optimality of the proposed constructions.
\end{remark}

	%	\textcolor{red}{To further illustrate the parameters of the constructed} LAZ sequence sets, specific periodic LAZ sequence set parameter values are presented in Tables \ref{table1} and \ref{table2} for $K=N$ and $N<K<2N-1$, respectively. Given that the optimality factor ${\rho}_{\mathrm{LAZ}}$ is a crucial metric for assessing the quality of LAZ sequence sets, we have also included it in this table. The numerical results indicate that the optimality factor ${\rho}_{\mathrm{LAZ}}$ of the constructed LAZ sequence sets approaches 1 asymptotically as $p$ increases. 
	\begin{table}[htp]
		\caption{Parameters of the proposed periodic LAZ sequence set when $K=N$}
		\label{table1}
		\begin{center}
			\begin{tabular}{|c|c|c|c|c|}
				\hline Set size  & Length  & Low ambiguity zone &  Maximum periodic AF & Optimality factor  \\
				$N$ & $N^2$ &  $\Pi$ &  magnitude $\theta_{\max}$ &  $\rho_{\mathrm{LAZ}}$ \\
				\hline 3 & 9 & $(-3,3) \times(-3,3)$ & 3 & 1.154701 \\
				\hline  5 & 25 & $(-5,5) \times(-5,5)$ & 5 & 1.095445 \\
				\hline  7 & 49 & $(-7,7) \times(-7,7)$ & 7 & 1.069045 \\
				\hline  11 & 121 & $(-11,11) \times(-11,11)$ & 11 & 1.044466 \\
				\hline  13 & 169 & $(-13,13) \times(-13,13)$ & 13 & 1.037749 \\
				\hline  17 & 289 & $(-17,17) \times(-17,17)$ & 17 & 1.028992 \\
				\hline  19 & 361 & $(-19,19) \times(-19,19)$ & 19 & 1.025978 \\
				\hline  23 & 529 & $(-23,23) \times(-23,23)$ & 23 & 1.021508 \\
				\hline  29 & 841 & $(-29,29) \times(-29,29)$ & 29 & 1.017095 \\
				%			\hline  31 & 961 & $(-31,31) \times(-31,31)$ & 31 & 1.016001 \\
				%			\hline  37 & 1369 & $(-37,37) \times(-37,37)$ & 37 & 1.013423 \\
				%			\hline 41 & 1681 &	$(-41,41) \times(-41,41)$& 41 & 1.012122 \\
				\hline
			\end{tabular}
		\end{center}
	\end{table}

	\begin{table}[htp]
		\caption{Parameters of the proposed periodic LAZ sequence set when $N<K<2N-1$}
		\label{table2}
		\begin{center}
			\begin{tabular}{|c|c|c|c|c|}
				\hline Set size  & Length  & Low ambiguity zone &  Maximum periodic AF & Optimality factor  \\
				$N$ & $NK$ &  $\Pi$ &  magnitude $\theta_{\max}$ &  $\rho_{\mathrm{LAZ}}$ \\
				\hline  7 & 77 & $(-7,7) \times(-5,5)$ & 11 & 1.498298 \\
				\hline  17 & 442 & $(-17,17) \times(-10,10)$ & 26 & 1.341383  \\
				\hline  31 & 1302 & $(-31,31) \times(-12,12)$ & 42 &1.235186 \\
				\hline 41 & 2132 &	$(-41,41) \times(-12,12)$& 52 & 1.190520  \\
				\hline  67 & 5360 & $(-67,67) \times(-14,14)$ & 80 & 1.142397 \\
				\hline 79 & 7347 &	$(-79,79) \times(-15,15)$& 93 & 1.130163\\
				\hline  89 & 9167 & $(-89,89) \times(-15,15)$ & 103 & 1.119777 \\
				\hline  101 & 11716 & $(-101,101) \times(-16,16)$ & 116 & 1.112300 \\
				\hline 127 & 18161 &	$(-127,127) \times(-17,17)$& 143 & 1.098079  \\
				\hline
			\end{tabular}
		\end{center}
	\end{table}

	%	\textcolor{red}{To further illustrate the parameters of the constructed} LAZ sequence sets, specific aperiodic LAZ sequence set parameter values are presented in Tables \ref{2} and \ref{3} for $K=N$ and $N<K<2N-1$, respectively. Given that the optimality factor $\hat{\rho}_{\mathrm{LAZ}}$ is a crucial metric for assessing the quality of LAZ sequence sets, we have also included it in this table. The numerical results indicate that the optimality factor $\hat{\rho}_{\mathrm{LAZ}}$ of the constructed LAZ sequence sets approaches 1 asymptotically as 
	%	$p$ increases. This confirms that the constructed LAZ sequence sets are asymptotically optimal, as predicted by Theorem \ref{Th4}.
	
	\begin{table}[htp]
		\caption{Parameters of the proposed aperiodic LAZ sequence set when $K=N$}
		\label{2}
		\begin{center}
			\begin{tabular}{|c|c|c|c|c|}
				\hline Set size  & Length  & Low ambiguity zone &  Maximum aperiodic AF & Optimality factor  \\
				$N$ & $N^2$ &  $\Pi$ &  magnitude $\hat{\theta}_{\max}$ &  $\hat{\rho}_{\mathrm{LAZ}}$ \\
				\hline 9 & 81 & $(-3,3) \times(-9,9)$ & 11 & 1.496217 \\
				\hline  15 & 225 & $(-3,3) \times(-15,15)$ & 17 & 1.381695 \\
				\hline 21 & 441 & $(-3,3) \times(-21,21)$ & 23 & 1.335227 \\
				\hline  25 & 625 & $(-5,5) \times(-25,25)$ & 29 & 1.296886 \\
				\hline  55 & 3025 & $(-5,5) \times(-55,55)$ & 59 & 1.198152 \\
				\hline  77 & 5929 & $(-7,7) \times(-77,77)$ & 83 & 1.163894 \\
				\hline  91 & 8281 & $(-7,7) \times(-91,91)$ & 97 & 1.150922 \\
				\hline  121 & 14641 & $(-11,11) \times(-121,121)$ & 131 & 1.135486 \\
				%			\hline  187 & 34969 & $(-11,11) \times(-187,187)$ & 197 & 1.104800 \\
				\hline  209 & 43681 & $(-11,11) \times(-209,209)$ & 219 & 1.098890 \\
				\hline 221 & 48841 & $(-13,13) \times(-221,221)$ & 233 & 1.097303 \\
				\hline
			\end{tabular}
		\end{center}
	\end{table}	
	
	\begin{table}[htp]
		\caption{Parameters of the proposed aperiodic LAZ sequence set when $N<K<2N-1$}
		\label{3}
		\begin{center}
			\begin{tabular}{|c|c|c|c|c|}
				\hline Set size  & Length  & Low ambiguity zone &  Maximum aperiodic AF & Optimality factor  \\
				$N$ & $NK$ &  $\Pi$ &  magnitude $\hat{\theta}_{\max}$ &  $\hat{\rho}_{\mathrm{LAZ}}$ \\
				\hline  25 & 950 & $(-5,5) \times(-14,14)$ & 42 & 2.016593 \\
				\hline  49 & 3479 & $(-7,7) \times(-23,23)$ & 77 &1.746188 \\
				\hline  121 & 20207 & $(-11,11) \times(-47,47)$ & 177 & 1.513314 \\
				\hline  169 & 38870 & $(-13,13) \times(-62,62)$ & 242 & 1.451976  \\
				\hline  289 & 110398 & $(-17,17) \times(-94,94)$ & 398 & 1.373160 \\
				\hline  529 & 359720 & $(-23,23) \times(-152,152)$ & 702 & 1.304136 \\
				\hline  961 & 1157044 & $(-31,31) \times(-244,244)$ & 1234 & 1.251085 \\
				\hline 1681 & 3466222 & $(-41,41) \times(-382,382)$ & 2102 & 1.211598 \\
				\hline  10201 & 120484011 & $(-101,101) \times(-1611,1611)$ & 11911 & 1.126801\\
				\hline 27889 & 878196721 & $(-127,127) \times(-3601,3601)$ & 31655 & 1.097300  \\
				\hline
			\end{tabular}
		\end{center}
	\end{table}

	\section{Concluding Remarks}\label{conclusion} 
	
	In this paper, we introduced a new class of Locally Perfect Nonlinear Functions (LPNFs) and proposed three new categories of periodic and aperiodic LAZ sequence sets with flexible parameters developed through interleaving techniques and LPNFs. Notably, two of these new classes of periodic and aperiodic LAZ sequence sets are asymptotically optimal with respect to the Ye-Zhou-Liu-Fan-Lei-Tang bounds. Additionally, the proposed LAZ sequences asymptotically satisfy the aperiodic LAZ Ye-Zhou-Liu-Fan-Lei-Tang bound. This is the first construction in literature, to satisfy the aperiodic LAZ Ye-Zhou-Liu-Fan-Lei-Tang bound. The constituent sequences within these LAZ sequence sets are cyclically distinct regarding both time shifts and phase shifts. The article includes several illustrations and explicit results presented in tables. Future research directions may include developing novel methodologies to design asymptotically optimal aperiodic LAZ sequence sets, particularly those with a prime length.

	%In this paper, we have proposed LPNFs and constructed three new classes of periodic and aperiodic LAZ sequence sets with flexible parameters using interleaving techniques and LPNFs. Two of the three new classes proposed periodic and aperiodic LAZ sequence sets are asymptotically optimal with respect to the Ye-Zhou-Liu-Fan-Lei-Tang bounds. The proposed LAZ sequences are the only sequences till date to asymptotically satisfy the aperiodic LAZ Ye-Zhou-Liu-Fan-Lei-Tang bound. The constituent sequences of the proposed LAZ sequence sets are cyclically distinct with respect to both time shift and phase shift.

	%	For the low ambiguity region $\Pi_1=(-p,p) \times (-N,N)$, $\Pi_2=(-p,p) \times (-K+N-1,K-N+1)$, $\Pi_3=(-p,p) \times (-K,K)$, with $N$ being an odd number, $K$ is a positive integer that satisfies $N<K<2N-1$ and $K\geq2N-1$, and $p$ being the smallest prime factor of $N$, the proposed parameters of the periodic LAZ sequence set are $(N,N^2,\Pi_1,N)$-LAZ, $(N,NK,\Pi_2,K)$-LAZ, and $(N,NK,\Pi_3,K)$-LAZ, respectively. The proposed aperiodic LAZ sequence set are $(N,N^2,\Pi_1,N+p-1)$-LAZ, $(N,NK,\Pi_2,K+p-1)$-LAZ, and$(N,NK,\Pi_3,K+p-1)$-LAZ, respectively.  
	%	The proposed sequence sets are a first of its kind to simultaneously satisfy both the periodic as well as aperiodic LAZ \textit{Ye-Zhou-Liu-Fan-Lei-Tang bounds}, asymptotically. The constituent sequences of the proposed LAZ sequence sets are cyclically distinct. %In the future, it would be
	%interesting, to explore asymptotically optimal aperiodic LAZ sequence sets when $N$ is a prime.

	\bibliographystyle{ieeetr}
	\bibliography{ref.bib}

\end{document}